\documentclass[lettersize,journal,]{IEEEtran}

\usepackage{xspace}
\xspaceaddexceptions{\%}
\xspaceremoveexception{-}

\usepackage[switch]{lineno}
\usepackage{graphicx}
\usepackage{subcaption} %
\usepackage{caption}
\usepackage{booktabs}
\usepackage{multirow}
\usepackage{tabularx}

\usepackage[hang,flushmargin]{footmisc} %

\usepackage{microtype}  %

\PassOptionsToPackage{%
  backend=bibtex,bibencoding=utf8,
  language=auto, %
  style=ieee,
  sorting=none, %
  minnames=1, %
  maxnames=1, %
  maxbibnames=10, %
  backref=false,%
  natbib=true, %
  doi=true,
  url=true
}{biblatex}
\usepackage{biblatex}
\usepackage{amsmath}
\usepackage{amssymb} %
\usepackage{amsfonts}
\usepackage{braket}  %
\usepackage[thinc]{esdiff}  %
\DeclareMathOperator{\Tr}{Tr}

\usepackage[compat=0.3]{yquant}  %
\usetikzlibrary{quotes}

\usepackage[inline]{enumitem}
\newlist{inlineenum}{enumerate*}{1}
\setlist[inlineenum]{label=(\alph*), itemjoin={{, }}, itemjoin*={{, and }}}

\usetikzlibrary{decorations.pathreplacing,calc}

\usepackage[framemethod=TikZ]{mdframed}
\newcommand{\anncmnt}[1]{%
\begin{mdframed}[backgroundcolor=gray!25,roundcorner=3pt,linewidth=1pt]%
  \noindent#1
\end{mdframed}%
}

\usepackage{algpseudocode}
\usepackage[linesnumbered,ruled,vlined]{algorithm2e}

\newcommand{\highlightline}[2]{%
  \noindent\colorbox{#1}{\parbox{\dimexpr\columnwidth-10\fboxsep}{#2}}%
}

\usepackage{url}

\usepackage{tikz}
\usepackage[framemethod=TikZ]{mdframed}
\usetikzlibrary{arrows}
\usetikzlibrary{shapes}

\usepackage[colorlinks]{hyperref} %
\usepackage{xcolor} %
\usepackage{colortbl}
\definecolor{gray}{rgb}{0.5,0.5,0.5}
\def\tmp#1#2#3{%
  \definecolor{Hy#1color}{#2}{#3}%
  \hypersetup{#1color=Hy#1color}}
\tmp{link}{HTML}{800006}
\tmp{cite}{HTML}{2E7E2A}
\tmp{file}{HTML}{131877}
\tmp{url} {HTML}{8A0087}
\tmp{menu}{HTML}{727500}
\tmp{run} {HTML}{137776}
\def\tmp#1#2{%
  \colorlet{Hy#1bordercolor}{Hy#1color#2}%
  \hypersetup{#1bordercolor=Hy#1bordercolor}}
\tmp{link}{!60!white}
\tmp{cite}{!60!white}
\tmp{file}{!60!white}
\tmp{url} {!60!white}
\tmp{menu}{!60!white}
\tmp{run} {!60!white}

\usepackage{cleveref}
\usepackage{balance}

\usepackage{csquotes}  %

\def\BibTeX{{\rm B\kern-.05em{\sc i\kern-.025em b}\kern-.08em
    T\kern-.1667em\lower.7ex\hbox{E}\kern-.125emX}}

\newcommand{\statezero}{\ket{00\cdots0}} 
\newcommand{\gate}[1]{\textsc{#1}} 

\newcommand{\cnot}{\gate{cx}}

\newcommand{\mcx}[1]{\textsc{c}\textsuperscript{#1}\textsc{x}}

\newcommand{\lIfElse}[3]{\lIf{#1}{#2 \textbf{else}~#3}}

\newcommand{\bug}{fault\xspace}

\newcommand{\buggy}{faulty\xspace}

\newcommand{\StatTest}{Statistical test\xspace}
\newcommand{\statTests}{statistical tests\xspace}
\newcommand{\StatTests}{Statistical tests\xspace}
\newcommand{\swapTest}{Swap test\xspace}
\newcommand{\statevectorTest}{Statevector test\xspace}
\newcommand{\inverseTest}{Inverse test\xspace}

\begin{document}

\title{On the Feasibility of Quantum Unit Testing}
\author{
    \IEEEauthorblockN{Andriy Miranskyy\IEEEauthorrefmark{1}, Jos\'{e} Campos\IEEEauthorrefmark{2}, Anila Mjeda\IEEEauthorrefmark{3}, Lei Zhang\IEEEauthorrefmark{4}, Ignacio Garc\'{i}a Rodr\'{i}guez de Guzm\'{a}n\IEEEauthorrefmark{5} \\ }
    \IEEEauthorblockA{\IEEEauthorrefmark{1}Department of Computer Science, Toronto Metropolitan University, Toronto, Canada \\}
    \IEEEauthorblockA{\IEEEauthorrefmark{2}Faculty of Engineering of University of Porto, Portugal; and LASIGE, Faculdade de Ci\^{e}ncias, Universidade de Lisboa, Portugal \\}
    \IEEEauthorblockA{\IEEEauthorrefmark{3}Department of Computer Science, Munster Technological University, Cork, Ireland\\ }
    \IEEEauthorblockA{\IEEEauthorrefmark{4}Department of Information Systems, University of Maryland, Baltimore County, Baltimore County, USA  \\}
    \IEEEauthorblockA{\IEEEauthorrefmark{5}Institute of Technologies and Information Systems, University of Castilla-La Mancha, Spain \\}
    \IEEEauthorblockA{avm@torontomu.ca, jcmc@fe.up.pt, anila.mjeda@mtu.ie, leizhang@umbc.edu, ignacio.grodriguez@uclm.es}
}

\maketitle

\begin{abstract}
The increasing complexity of quantum software presents significant challenges for software verification and validation, particularly in the context of unit testing. This work presents a comprehensive study on quantum-centric unit tests, comparing traditional statistical approaches with tests specifically designed for quantum circuits. These include tests that run only on a classical computer, such as the \statevectorTest, as well as those executable on quantum hardware, such as the \swapTest and the novel \inverseTest.
Through %
an empirical study and detailed analysis on
1,796,880 mutated quantum circuits,
we investigate
\begin{inlineenum}
\item each test's ability to detect subtle discrepancies between the expected and actual states of a quantum circuit
\item the number of measurements required to achieve high reliability.
\end{inlineenum}
The results demonstrate that quantum-centric tests, particularly the \statevectorTest and the \inverseTest, provide clear advantages in terms of precision and efficiency, reducing both false positives and false negatives compared to statistical tests. This work contributes to the development of more robust and scalable strategies for testing quantum software, supporting the future adoption of fault-tolerant quantum computers and promoting more reliable practices in quantum software engineering.
\end{abstract}

\section{Introduction}\label{sec:intro}

The increasing maturity of quantum computing technologies has driven the emergence of more sophisticated quantum algorithms and programs. As these programs become more complex, ensuring their correctness and reliability becomes critical, particularly given the non-intuitive nature of quantum mechanics~\cite{nielsen_chuang_2010}. Traditional software engineering practices, such as unit testing, have been extensively used to validate classical programs~\cite{daka2014survey}. However, the direct adaptation of these techniques to quantum software faces unique challenges~\cite{miranskyy2019testing, miranskyy2020quantum,miranskyy2021testing,piattini2020Towards,talaveraManifesto2020,zhao2021qselandscapes,ramalho2024testing}, including the exponential growth of the state space and the probabilistic outcomes inherent to quantum measurements.

To address these challenges, researchers have proposed various approaches to
validate and verify quantum programs (see \cite{ramalho2024testing,murillo2024challenges} for a full review on this topic).  The three most popular ones are: the \StatTests, the \swapTest~\cite{barenco1997,buhrman2001}, and the \statevectorTest.

\StatTests compare the distributions of expected and actual outputs to identify discrepancies.
While effective in certain scenarios, statistical tests often require a large number of \emph{measurements} (also known as \emph{shots}) and may produce false positives or negatives, which can hinder practical applicability.
The \swapTest estimates the overlap between the expected and actual quantum states by using an ancillary qubit and a controlled-SWAP operation.  By design, the \swapTest does not produce false positives, but can yield false negatives and may require many shots.
The \statevectorTest compares the actual and expected \textit{quantum state vectors} (hereafter abbreviated as \textit{state vectors}) using classical simulation of the quantum circuit, offering precise and accurate results.  However, it runs only on a classical computer and scales exponentially with the width of the quantum register.

Consider, as a motivational example, a trivial quantum circuit where a single qubit is initialized in the $\ket{0}$ state and placed into superposition using a Hadamard gate:
$$H = \frac{1}{\sqrt{2}} \begin{bmatrix} 1 & 1 \\ 1 & -1 \end{bmatrix} \approx \begin{bmatrix} 0.7071 & 0.7071 \\ 0.7071 & -0.7071 \end{bmatrix}. $$
Applying this gate to $\ket{0}$ yields the expected state vector:
$$\ket{\phi_{\text{exp}}} = H \ket{0} = \left[ \frac{1}{\sqrt{2}}, \frac{1}{\sqrt{2}} \right]^\intercal \approx \left[ 0.7071, 0.7071 \right]^\intercal, $$
where $\intercal$ denotes the transpose.
Now, suppose an error is introduced into the Hadamard gate, resulting in a slightly perturbed unitary operation:
$$\hat H \approx \begin{bmatrix} 0.7066 & 0.7076 \\ 0.7076 & -0.7066 \end{bmatrix}. $$
The resulting \buggy state\footnote{The root cause of this small deviation could lie at various levels of the software or hardware stack: e.g., a bug in the circuit defined by the programmer, a defect in the gate implementation, an issue in the transpiler, or a hardware fault. However, from the perspective of our example, the specific cause is irrelevant; we simply observe that $\ket{\phi_\text{exp}} \neq \ket{\phi_\text{act}}$.} is therefore:
$$\ket{\phi_{\text{bug}}} = \hat H \ket{0} \approx \left[ 0.7066, 0.7077 \right]^\intercal, $$
which deviates from the expected one:
$$\ket{\phi_{\text{bug}}} - \ket{\phi_{\text{exp}}} \approx \left[ -0.0005, 0,0006 \right]^\intercal.$$

{\small\begin{table*}[tb]
\centering
\caption{Performance of each test in the motivational example. \\
{\small 
For each test, we report the number of true positives (TP), number of true negatives (TN), number of false positives (FP), number of false negatives (FN), accuracy (Acc.), precision (Prec.), recall (Rec.), and F$_1$ score (F$_1$).  MC stands for ``Monte Carlo''. Best values are reported in \textbf{bold}. For all columns, with the exception of FP and FN, higher values indicate better performance.}\label{tbl:motivational_example_stats}}
\vspace{-1em}
\begin{tabular}{@{}l|rrrrrrrr|rrrrrrrr@{}}
\toprule
Test Name
& \multicolumn{8}{c|}{$p < 0.05$} 
& \multicolumn{8}{c}{$p < 0.01$} \\
\cmidrule(lr){2-9} \cmidrule(lr){10-17}
& TP & TN & FP & FN & Acc. & Prec. & Rec. & F$_1$ 
& TP & TN & FP & FN & Acc. & Prec. & Rec. & F$_1$ \\

\midrule
\rowcolor{gray!25}
\multicolumn{17}{c}{\textbf{\textit{Number of shots = $10^5$, as in recent studies on quantum software testing~\cite{UsandizagaMutation2025}}}} \\
\midrule
$\chi^2$ test     & 79  & 127 & 73  & 121 & 0.515 & 0.520 & 0.395 & 0.449 & 20  & 182 & 18  & 180 & 0.505 & 0.526 & 0.100 & 0.168 \\
G-test            & 79  & 127 & 73  & 121 & 0.515 & 0.520 & 0.395 & 0.449 & 20  & 182 & 18  & 180 & 0.505 & 0.526 & 0.100 & 0.168 \\
Multinomial test  & \textbf{200} & 0   & 200 & \textbf{0}   & 0.500 & 0.500 & \textbf{1.000} & 0.667 & \textbf{200} & 0   & 200 & \textbf{0}   & 0.500 & 0.500 & \textbf{1.000} & 0.667 \\
MC $\chi^2$ test  & 63  & 143 & 57  & 137 & 0.515 & 0.525 & 0.315 & 0.394 & 15  & 187 & 13  & 185 & 0.505 & 0.536 & 0.075 & 0.132 \\
MC G-test         & 64  & 136 & 64  & 136 & 0.500 & 0.500 & 0.320 & 0.390 & 16  & 187 & 13  & 184 & 0.508 & 0.552 & 0.080 & 0.140 \\
MC Multinomial test & 64  & 135 & 65  & 136 & 0.498 & 0.496 & 0.320 & 0.389 & 17  & 186 & 14  & 183 & 0.508 & 0.548 & 0.085 & 0.147 \\
\swapTest     & 5   & \textbf{200} & \textbf{0}   & 195 & 0.513 & \textbf{1.000} & 0.025 & 0.049 & 5   & \textbf{200} & \textbf{0}   & 195 & 0.513 & \textbf{1.000} & 0.025 & 0.049 \\
\statevectorTest$^*$    & \textbf{200} & \textbf{200} & \textbf{0} & \textbf{0} & \textbf{1.000} & \textbf{1.000} & \textbf{1.000} & \textbf{1.000}  & \textbf{200} & \textbf{200} & \textbf{0} & \textbf{0} & \textbf{1.000} & \textbf{1.000} & \textbf{1.000} & \textbf{1.000} \\
\inverseTest  & 9   & \textbf{200} & \textbf{0}   & 191 & 0.523 & \textbf{1.000} & 0.045 & 0.086 & 9   & \textbf{200} & \textbf{0}   & 191 & 0.523 & \textbf{1.000} & 0.045 & 0.086 \\

\midrule
\rowcolor{gray!25}
\multicolumn{17}{c}{\textbf{\textit{Number of shots = $10^7$, i.e., above the theoretical
prediction values based on \Cref{eq:estimate_of_s} using the Quantum Chernoff Bound~\cite{audenaert2007discriminating,nussbaum2009chernoff}}}} \\
\midrule
$\chi^2$ test     & \textbf{200} & 120 &  80 &  \textbf{0} & 0.800 & 0.714 & \textbf{1.000} & 0.833  & 197 & 189 &  11 &  3 & 0.965 & 0.947 & 0.985 & 0.966 \\
G-test            & \textbf{200} & 120 &  80 &  \textbf{0} & 0.800 & 0.714 & \textbf{1.000} & 0.833  & 197 & 189 &  11 &  3 & 0.965 & 0.947 & 0.985 & 0.966 \\
Multinomial test  & \textbf{200} &   0 & 200 &  \textbf{0} & 0.500 & 0.500 & \textbf{1.000} & 0.667  & \textbf{200} &   0 & 200 &  \textbf{0} & 0.500 & 0.500 & \textbf{1.000} & 0.667 \\
MC $\chi^2$ test  & \textbf{200} & 128 &  72 &  \textbf{0} & 0.820 & 0.735 & \textbf{1.000} & 0.847  & 196 & 188 &  12 &  4 & 0.960 & 0.942 & 0.980 & 0.961 \\
MC G-test         & \textbf{200} & 123 &  77 &  \textbf{0} & 0.808 & 0.722 & \textbf{1.000} & 0.839  & 196 & 191 &   9 &  4 & 0.968 & 0.956 & 0.980 & 0.968 \\
MC Multinomial test & \textbf{200} & 126 &  74 &  \textbf{0} & 0.815 & 0.730 & \textbf{1.000} & 0.844  & 197 & 186 &  14 &  3 & 0.958 & 0.934 & 0.985 & 0.959 \\
\swapTest     & 191 & \textbf{200} & \textbf{0} &  9 & 0.978 & \textbf{1.000} & 0.955 & 0.977  & 191 & \textbf{200} & \textbf{0} &  9 & 0.978 & \textbf{1.000} & 0.955 & 0.977 \\
\statevectorTest$^*$    & \textbf{200} & \textbf{200} & \textbf{0} & \textbf{0} & \textbf{1.000} & \textbf{1.000} & \textbf{1.000} & \textbf{1.000}  & \textbf{200} & \textbf{200} & \textbf{0} & \textbf{0} & \textbf{1.000} & \textbf{1.000} & \textbf{1.000} & \textbf{1.000} \\
\inverseTest  & \textbf{200} & \textbf{200} & \textbf{0} & \textbf{0} & \textbf{1.000} & \textbf{1.000} & \textbf{1.000} & \textbf{1.000}  & \textbf{200} & \textbf{200} & \textbf{0} & \textbf{0} & \textbf{1.000} & \textbf{1.000} & \textbf{1.000} & \textbf{1.000} \\

\bottomrule
\multicolumn{17}{l}{$^{*}$Note that the \statevectorTest is independent of the number of shots. It is included in both parts of the table to facilitate comparison. }
\end{tabular}
\end{table*}}

On one hand, if we run the fault-free circuit on a noise-free simulator and measure its single-qubit register, we should observe the outcomes $0$ and $1$ with equal probability of $50.0\%$ each~---~essentially simulating an unbiased coin flip.  On the other hand, if we measure the single-qubit register in the \buggy circuit, the probability of observing $0$ is $\approx$$49.9\%$ and the probability of observing $1$ is  $\approx$$50.1\%$, indicating a slightly biased coin flip.
Given the small differences %
in the probabilities of the outcomes in the \bug free vs.\ \buggy circuit, we could ask:
\begin{center}\emph{%
Would the previously proposed \StatTests, the \swapTest,\\
or the \statevectorTest be able to reliably distinguish\\
between the expected and the \buggy state?
}\end{center}

To answer this question, we ran 200 independent repetitions of six statistical tests. %
\begin{enumerate}[leftmargin=*]
  \item The $\chi^2$ test~\cite{pearson1900x,mcdonald2014handbook}, which has been used in several other works on quantum software testing~\cite{wang2022qusbt,wang2023qucat,UsandizagaMutation2025}.
\end{enumerate}
Two other that, although powerful, have not been explored in the context of quantum software testing, namely:
\begin{enumerate}[leftmargin=*]\setcounter{enumi}{1}
  \item The G-test~\cite{mcdonald2014handbook},
  \item The Multinomial test~\cite{mcdonald2014handbook}.
\end{enumerate}
And their respective Monte Carlo (MC) variants:
\begin{enumerate}[leftmargin=*]\setcounter{enumi}{3}
  \item MC $\chi^2$ test,
  \item MC G-test,
  \item MC Multinomial test.
\end{enumerate}
Additionally, we also considered the \swapTest and the \statevectorTest.  As for the number of shots we set it to $10^5$ (as recently suggested by~\citet{UsandizagaMutation2025}) and $10^7$ (based on the theoretical estimation\footnote{\Cref{eq:estimate_of_s} indicates that our \inverseTest (later presented in \Cref{sec:methodology_quite_measurement_test}) can distinguish the different between the states with a $p$-value of $0.05$ using using $5.4 \times 10^6$ shots, and with a $p$-value of $0.01$ using $8.3 \times 10^6$ shots. The chosen value of $10^7$ shots accommodates both cases.} of \Cref{eq:estimate_of_s} using the Quantum Chernoff Bound~\cite{audenaert2007discriminating,nussbaum2009chernoff}).  While the latter might be high for practical applications, it provides a conceptual upper bound for analysis. Later in the paper, we explore how few shots are actually necessary for reliable decision-making.  Additionally, the statistical tests were configured with two $p$-values commonly used in the literature: $0.05$ (e.g.,~\cite{wang2022qusbt}) and $0.01$ (e.g.,~\cite{wang2021quito,wang2023qucat,UsandizagaMutation2025}).

Then, we defined two scenarios:
\begin{enumerate}[leftmargin=*]
  \item \textit{Negative case}: The actual and expected states are identical accordingly to the test: $\ket{\phi_\text{act}} = \ket{\phi_\text{exp}} = H\ket{0}$. In other words, there is no \bug.

  \item \textit{Positive case}: The actual and \buggy states differ slightly accordingly to the test: $\ket{\phi_\text{act}} = \ket{\phi_\text{bug}}  = \hat H\ket{0}$ and $\ket{\phi_\text{exp}} = H\ket{0}$. Here, a \bug is present.
\end{enumerate}
And framed the answer to our question as a binary classification problem.  When $\ket{\phi_\text{act}} = \ket{\phi_\text{exp}}$, the ideal outcome is a \textit{true negative}~---~the test should not report a failure. When the states differ, i.e., $\ket{\phi_\text{act}} \neq \ket{\phi_\text{exp}}$, the correct outcome is a \textit{true positive}~---~the test should report a failure.
Misclassifications fall into two categories: Type I error, \textit{false positives} (reporting a failure when the states are equal) and Type II error, \textit{false negatives} (failing to report a failure when the states differ). These errors are analogous to flaky tests, i.e., tests that inconsistently pass or fail under the same conditions~\cite{zhang2023identifying,zhang2024automated,sivaloganathan2024automating}.

The results are summarized in \Cref{tbl:motivational_example_stats}, and the key observations are as follows.

First, consider the case of $10^5$ shots. When the number of shots is significantly lower than the theoretically estimated value, all tests perform poorly. Interestingly, the \swapTest and \inverseTest occasionally yield correct results by chance, indicating that even a small number of measurements can sometimes succeed. However, while $10^5$ may seem large, achieving consistent and reliable results requires more shots. Below, we analyze what happens when the number of shots exceeds the estimated recommended value, specifically $10^7$.

Statistical tests produce both false positives and false negatives. False positives are particularly problematic, as they indicate failure even when the code is correct, frustrating for developers.
Adjusting the $p$-value threshold in statistical tests trades off false positives and false negatives. For example, lowering the threshold from $0.05$ to $0.01$ reduced false positives from $80$ to $11$ in the $\chi^2$ test, but increased false negatives from $0$ to $3$.
The Multinomial test is highly sensitive and ``latches'' onto positive cases. This issue is mitigated by using a Monte Carlo version, which samples the expected distribution a number of times\footnote{We set the number of repetitions in the Monte Carlo statistical test to 1,000; more details in Section~\ref{sec:mc_reps}.} to improve robustness. Overall, the Monte Carlo versions of all three statistical tests show slightly better performance in terms of Accuracy, Precision, Recall, and F$_1$ Score.

The \swapTest~\cite{barenco1997,buhrman2001} has been designed to produce no false positives in ideal, noise-free circuits (which is our case), i.e., correctly report no difference when the actual and expected states are equal.
However, when the states differ, the test may yield false negatives if the number of shots is insufficient.  Although the \swapTest reports nine false negatives vs.\ zero reported by any statistical test, it achieves a higher Accuracy, Precision, and F$_1$ Score than of the statistical test.

The \statevectorTest computes and compares the actual and expected state vectors on a classical computer, yielding 100\% accuracy and precision, as expected. It is independent of the number of shots because the state vectors are obtained by multiplying the matrix representations of the circuit gates. However, the cost of computing state vectors grows exponentially with the number of qubits.

\textbf{Our work.}
To overcome the limitations of the \StatTests, the \swapTest, and the \statevectorTest, we propose a novel test, the \emph{\inverseTest}.  It is based on circuit reversion and encodes the expected state behavior directly into the quantum circuit, enabling pass/fail outcomes through measurements. %
Unlike the \statTests but similar to the \swapTest, the \inverseTest operates on a fixed measurement rule: a particular observed output directly indicates test failure.  If that output is not observed, more shots may be needed.  This binary property makes it conceptually simpler to analyze.
Additionally, and similar to the \swapTest, the \inverseTest is designed to produce no false positives.

Revisiting the motivational example, the \inverseTest is on par with the \statevectorTest, outperforming the \StatTests and the \swapTest across all key performance metrics (Accuracy, Precision, Recall, and F$_1$ Score). This suggests that the \inverseTest has higher statistical power and may require fewer shots to detect subtle differences, making it more computationally efficient. %
However, \emph{does this advantage persist in other and perhaps more complex quantum circuits?}

To answer this overarching question, we design and conduct an experimental evaluation where the actual output of 1,796,880 quantum circuits intentionally differs from its expected output.  We achieve this by mutating 10,000 randomly generated quantum circuits and 85 circuits from the MQT Bench~\cite{quetschlich2023mqtbench}.
We then
\begin{inlineenum}
\item assess how many shots each test requires to reliably detect the discrepancy between each mutant and its original version
\item analyze tests' false positive and false negative rates
\item discuss tests' overall resource efficiency.
\end{inlineenum}
The results show that the \statevectorTest and the \inverseTest are the strongest contenders.

Our \textbf{contributions} are as follows:
\begin{itemize}[leftmargin=*]
  \item[\small{$\bigstar$}] \textbf{Formal definition of a quantum unit test}, a test harness inspired by classical software testing, that precisely describes what constitutes a unit test for quantum circuits. (\Cref{sec:qut_formal})

  \item[\small{$\bigstar$}] \textbf{Implementation of \StatTests, the \swapTest, and \statevectorTest} as a quantum unit test. (\Cref{sec:methodology_stats_tests,sec:methodology_swap_test,sec:methodology_quite_statevector_test})

  \item[\small{$\bigstar$}] \textbf{Definition and implementation of a novel test, the \inverseTest}, as a quantum unit test. %
  (\Cref{sec:methodology_quite_measurement_test})

  \item[\small{$\bigstar$}] \textbf{Evidence of tests' efficiency.}
  We analyze the theoretical time complexity of the quantum unit tests under study, comparing their performance across different setups. (\Cref{sec:complexity_analysis})   

  \item[\small{$\bigstar$}] \textbf{Evidence of tests' effectiveness.}
  We empirically evaluate across 1,796,880 pairs of original-mutant circuits (2.5 times more mutants than the previous largest empirical study on quantum mutation testing~\cite{UsandizagaMutation2025}, 723,079) and nine quantum unit tests, and show that \statevectorTest and the \inverseTest reliably distinguish between quantum states that are slightly different.
  (\Cref{sec:case_study})
\end{itemize}

\section{Background and Related work}\label{sec:related_literature}

\noindent\textbf{Quantum assertions vs.\ Quantum testing.}
In classical computing, a runtime assertion placed in the program's source code is a predicate used to verify the correctness of a program during its execution.  If the assertion passes, the program continues; if it fails, the program typically\footnote{Runtime assertions are, by default, disabled in Java but enabled in Python, for example.} terminates.  %

Significant work has been conducted on quantum assertions~\cite{StatisticalAssertions,witharana2023quassert,zhou2019quantum,DynamicAsseertions,li2022exploiting,patel2021qraft,chen2023vanqira,de2024quantumunit,rovara2024automatically,rovara2024framework,li2020projection,huang2019statistical,liu2021systematic} and we refer the reader to~\cite{murillo2024challenges,ramalho2024testing,li2022exploiting} for a full review.
Consider, for example, the statistical assertions~\cite{huang2019statistical}, that use statistical tests, such as the $\chi^2$ test, to compare multiple measurement outcomes of a qubit versus its expected state. These methods are destructive: once a quantum measurement is performed, the system collapses, and we cannot add further operations to the measured qubit(s).  This behavior diverges significantly from the classical understanding of a runtime assertion.  For clarity and consistency, in this paper, we refer to such checks as \emph{statistical test} rather than a \emph{statistical assertion} and we encapsulate them within the structure of a quantum unit test (\Cref{sec:methodology_stats_tests}).

On the other hand, consider the \swapTest, which involves measuring an auxiliary qubit while leaving the data qubits unchanged (provided that the test passes).  This non-destructive behavior allows further operations to be applied to the data qubits after the measurement.  As such, the Swap Test more closely aligns with a classical runtime assertion: we can perform a mid-circuit measurement and decide to terminate the program if we measure the ``0" value (i.e., assertion passes) on the auxiliary qubit or continue otherwise.
Thus, although we use the \emph{\swapTest} as a quantum unit test %
(\Cref{sec:methodology_swap_test}), it can also be used as a legitimate quantum runtime assertion (unlike the destructive statistical assertions).

\smallskip\noindent\textbf{Output probability oracle vs.\ Quantum testing.}
Statistical tests such $\chi^2$ have been used as oracles to assess whether the distribution of measurement outcomes of a quantum circuit is similar to the expected distribution~\cite{wang2021quito,wang2022qusbt,wang2023qucat,QuraTest,UsandizagaMutation2025,QuCheck}.  We leverage this knowledge and encapsulate this oracle in a quantum unit test (\Cref{sec:methodology_stats_tests}).

\smallskip\noindent\textbf{Number of measurements (shots).}
Qiskit~\cite{Qiskit}, one of the most popular quantum frameworks~\cite{FERREIRA2025103217}, sets the number of shots to 1,024, by default.
Others, in quantum software testing, have used a fixed number of shots in their empirical evaluations.  For example, \citet{wang2021quito} used a single shot\footnote{Number of shots performed by QUITO~\cite{wang2021quito} \url{https://github.com/Simula-COMPLEX/quito/blob/main/Quito_CoverageRunning/quito.py\#L141}, accessed July 2025.}, \citet{QuraTest} used 10,000 shots\footnote{Number of shots performed by QuraTest~\cite{QuraTest} \url{https://github.com/ToolmanInside/quratest/blob/main/oracle.py\#L151}, accessed July 2025.}, and \citet{UsandizagaMutation2025} used 100,000 shots.
To the best of our knowledge, only one study on quantum software testing has explicitly considered multiple shot counts: \citet{QuCheck} evaluated 12, 25, 50, 100, 200, 400, 800, 1,600, and 3,200 shots. %

In contrast, we estimate the number of shots using the Quantum Chernoff Bound approach~\cite{audenaert2007discriminating,nussbaum2009chernoff} \emph{for each circuit}, rather than relying on predefined values.

\smallskip\noindent\textbf{Quantum test case circuit vs.\ Quantum unit testing.}
As noted in~\cite[Fig.~14]{ramalho2024testing} and to the best of our knowledge, there %
is no well-defined definition of what constitutes a quantum-centered unit test for quantum programs.  Nevertheless, the closest approach to ours, named QTCC (Quantum Test Case Circuit)~\cite{garciaDelAmo2022testcase}, formalizes the idea of a quantum test case and provides a pedagogical circuit-based example. As such, it can be seen as a conceptual precursor to our work, proposing a quantum circuit that embeds:
\begin{inlineenum}
    \item the quantum circuit under test
    \item the expected value
    \item the oracle that verifies the equivalence between the latter and the value obtained when executing the circuit under test.
\end{inlineenum}

\section{Quantum unit test}\label{sec:qut_formalism}
In this section, we formalize the notion of a quantum unit test by adapting key principles from classical unit testing to the quantum computing context. \Cref{sec:qut_formal} introduces the formal definition of a quantum unit test, detailing its three essential components: the input, the program under test, and the oracle used to verify correctness. \Cref{sec:methodology_qut} follows with concrete implementations of four representative quantum unit tests (Statistical, Swap, Statevector, and Inverse) each structured around the classical Arrange-Act-Assert paradigm~\cite{ArrangeActAssert} and adapted to account for the unique properties of quantum circuits.

\subsection{Formal definition of a quantum unit test}\label{sec:qut_formal}

While some concepts of testing software for classical computers apply to quantum software, many others do not~\cite{miranskyy2019testing, miranskyy2020quantum, miranskyy2021testing}. In the case of unit testing, the core components remain the same~\cite{garciaDelAmo2022testcase}:
\begin{enumerate*}
\item an input,
\item the program under test, and
\item the oracle, i.e., a mechanism to determine whether the actual output matches the expected output.
\end{enumerate*}
In this section, we present the quantum formalism for these three components, which will support the implementation of quantum unit tests in \Cref{sec:methodology_qut}.

\subsubsection{Input} 

The \textit{input state} is defined as $\ket{\psi_I}$. Assume this state is described by $n$ qubits, with 
$$\ket{\psi_0} = \statezero = \ket{0}^{\otimes n}$$
as the default starting state in modern quantum computers. Also, assume that the input state can be obtained by applying a unitary operator $W$ to $\ket{\psi_0}$, i.e., 
$$\ket{\psi_I} = W \ket{\psi_0}.$$
Further assume that $W$ is correctly constructed and free from implementation defects.

When the input state is $\statezero$, $W$ is an identity matrix. For more complex input states, $W$ must be defined explicitly (e.g., as a state vector or unitary operator) to generate the corresponding gate sequence to transform $\ket{\psi_0}$ into $\ket{\psi_I}$.

If a circuit is partitioned into multiple subroutines and the gates preceding the subroutine under test are known to be correct, the circuit can compute $\ket{\psi_I}$. However, for better modularization and adherence to software engineering practices, generating the state independently (e.g., using a state vector to gate sequence converter) is preferred.

\subsubsection{Program under test}

The program under test is represented by a unitary operator $U$. 
It can be the sequence of quantum gates of a complete quantum circuit,
a sub-sequence of gates extracted from the quantum circuit,
or an actual subroutine in the OpenQASM~3.0~\cite{cross2022openqasm} sense.
In this paper, we consider the program under test $U$ as the quantum circuit of a quantum program, and we may use the terms quantum circuit or circuit interchangeably, in which case we imply the former. %

\subsubsection{Test oracle}
\paragraph{Expected State vs. Actual State}
The expected state $\ket{\psi_E}$ is a complete theoretical description of the quantum state that the circuit is intended to produce. It is represented as a state vector containing all the amplitudes (with both magnitude and phase) for every basis state.
For example, if we design a circuit to create an entangled state, $\ket{\psi_E}$ captures the exact superposition that should exist.
Knowing $\ket{\psi_E}$ allows for a precise, element-by-element comparison against the actual state $\ket{\psi_A}$.

The actual state $\ket{\psi_A}$ represents the outcome of applying the input $\ket{\psi_I}$ to $U$ and it is defined as
$$\ket{\psi_A} = U \ket{\psi_I}.$$

\paragraph{Unknown Expected State}

Note that $\ket{\psi_E}$ must be clearly defined (e.g., as a known state vector) in order to verify the correctness of $\ket{\psi_A}$. As in classical unit testing, knowing the expected output is crucial to confirm the program's correctness. A \textit{core limitation} of our approach arises if the expected output for a given program or subroutine is unknown: in that case, one cannot perform a meaningful test. An attempt to alleviate this problem was proposed by \citet{9407042}.  They proposed an approximate test oracle that assesses whether the actual state of a circuit is within a given set of states or a superposition of the states in the set.

That said, this challenge mirrors the age-old problem in classical testing where a test oracle does not always exists to provide a definitive expected output~\cite{oracleProblemInClassic,5298471,OLIVEIRA2014113,PEZZE20141}. In classical software testing, when an expected output is unavailable, empirical methods are often employed. For instance, differential testing can provide insight by executing multiple implementations of the same system, and comparing their results~\cite{mckeeman1998differential,7965305,8804465,Evans2007}. In simulation, discrepancies between different implementations may indicate potential errors, allowing for indirect verification. Additionally, in some cases, classically computed probabilities or approximations can serve as a reference to compare against $\ket{\psi_A}$ enabling some level of validation even when the exact expected state is unknown. Equivalent approaches can also be applied or further developed for quantum testing;  however, exploring these methods is outside the scope of this paper.

\paragraph{Comparison of Expected and Actual States}
The test oracle provides a specification of the expected state $\ket{\psi_E}$, representing the correct output of a quantum program for a given input. A separate \emph{mechanism} is then required to compare the actual output state $\ket{\psi_A}$, produced by the circuit under test, against this expected state. If they match (i.e., $\ket{\psi_E} = \ket{\psi_A}$), the test passes; if they differ (i.e., $\ket{\psi_E} \neq \ket{\psi_A}$), the test fails. The next section discusses various methods for performing this comparison.

\subsection{Implementation of quantum unit tests}\label{sec:methodology_qut}
\makeatletter
\DeclareRobustCommand\rvdots{%
\vbox{%
\baselineskip4\p@\lineskiplimit\z@%
\kern-\p@%
\hbox{.}\hbox{.}\hbox{.}%
}%
}
In this section, we describe, as quantum unit tests, the implementation of the \StatTests, the modified \swapTest, the \statevectorTest, and the novel \inverseTest in \Cref{sec:methodology_stats_tests,sec:methodology_swap_test,sec:methodology_quite_statevector_test,sec:methodology_quite_measurement_test}, respectively.
To enhance the readability and maintainability of quantum unit tests, each test is structured in three distinct phases as suggested by others for unit tests in the classical realm~\cite{ArrangeActAssert}.
The {\colorbox{yellow!30}{\textbf{Arrange}}} phase is responsible for initializing a quantum testing circuit with an input state ($\statezero$ by default) and for appending the quantum program under test to the testing circuit.
The {\colorbox{blue!30}{\textbf{Act}}} phase runs the testing circuit by evolving the quantum state through the unitary operations in the circuit.
And finally, the {\colorbox{red!30}{\textbf{Assert}}} phase checks whether the actual states matches the expected state.

\subsubsection{Statistical tests}\label{sec:methodology_stats_tests}
Commonly used to measure discrepancies between actual and expected outcomes. The process involves performing $S$ shots and recording the measured bitstrings\footnote{We assume that all $n$ qubits in the quantum register are measured. If only a subset is measured, the same principles apply.} from the quantum register implementing $\ket{\psi_A}$. The resulting distribution is then compared with the theoretical distribution derived from $\ket{\psi_E}$.
Statistical tests, by design, are susceptible to false positives and false negatives errors due to the inherent limitations of frequentist statistics~\cite[Sec. 8.3]{casella2002statistical}. The likelihood of such errors depends on the specific test, sample size, and the chosen $p$-value threshold.

\begin{figure}[t]
    \centering
    \begin{tikzpicture}
      \begin{yquant}
        qubit {$\ket{q_1} = \ket{0}$} q;
        qubit {$\ket{q_2} = \ket{0}$} q[+1];
        qubit {$\rvdots$} q[+1]; discard q[2];
        qubit {$\ket{q_n} = \ket{0}$} q[+1];
        box {$W$} (q);
        box {$U$} (q);
        measure q[0,1,3];
      \end{yquant}
    \end{tikzpicture}
    \caption{Quantum testing circuit for the \StatTests.} %
    \label{fig:stats_test_circuit}
\end{figure}
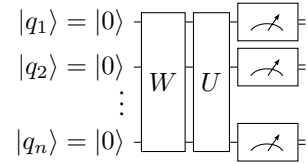
\begin{algorithm}[tb]
\SetKwComment{Header}{}{}
\SetAlgoLined
\SetKwInOut{Parameter}{Parameter}
\Parameter{$W$ as the input state}
\Parameter{$U$ as the software under test}
\Parameter{$\ket{\psi_E}$ as the expected output state}
\Parameter{$S$ as the number of shots}
\Parameter{$p_t$ as a significance threshold for statistical test}
\Parameter{$\textit{statTest}$ as the statistical test function, i.e., $\chi^2$ test or G-test}
\SetKwInOut{Input}{input}
\KwResult{Test result: Pass ($U$ behaves as expected) or Fail ($U$ does not behave as expected)}
\Header{\highlightline{yellow!30}{\textbf{Arrange}}}
Create an empty quantum circuit with a $n$-qubit register\;
Append $W$ then $U$ to the circuit\;
Add measurements to all qubits\;
\Header{/* At this point, the circuit looks like the one depicted in \Cref{fig:stats_test_circuit} */}
\Header{\highlightline{blue!30}{\textbf{Act}}}
Run the circuit $S$ times\;
$V \leftarrow$ Collect measurement outcomes\;
\Header{\highlightline{red!30}{\textbf{Assert}}}
$V_E \leftarrow$ Generate expected counts from $\ket{\psi_E}$\;
$p_s \leftarrow$ \textit{statTest}$(V, V_E)$\;
\lIfElse{$p_s \ge p_t$}{\Return Pass}{\Return Fail}%
\caption{\StatTest.} %
\label{alg:stats_test}
\end{algorithm}
\begin{algorithm}[tb]
\SetKwComment{Header}{}{}
\SetAlgoLined
\SetKwInOut{Parameter}{Parameter}
\Parameter{$W$ as the input state}
\Parameter{$U$ as the software under test}
\Parameter{$\ket{\psi_E}$ as the expected output state}
\Parameter{$S$ as the number of shots}
\Parameter{$p_t$ as a significance threshold for statistical test}
\Parameter{$\textit{statTest}$ as the statistical test function, e.g., $\chi^2$}
\Parameter{$M$ as a number of Monte Carlo repetitions} %
\SetKwInOut{Input}{input}
\KwResult{Test result: Pass ($U$ behaves as expected) or Fail ($U$ does not behave as expected)}
\Header{\highlightline{yellow!30}{\textbf{Arrange}}}
Create an empty quantum circuit with a $n$-qubit register\;
Append $W$ then $U$ to the circuit\;
Add measurements to all qubits\;
\Header{/* At this point, the circuit looks like the one depicted in \Cref{fig:stats_test_circuit} */}
\Header{\highlightline{blue!30}{\textbf{Act}}}
Run the circuit $S$ times\;
$V \leftarrow$ Collect measurement outcomes\;
\Header{\highlightline{red!30}{\textbf{Assert}}}
$V_E \leftarrow$ Generate expected counts from $\ket{\psi_E}$\;
$s_e \leftarrow$ \textit{statTest}$(V, V_E)$\;
$c \leftarrow 0$ \tcp*{Counter for Monte Carlo samples $\geq s_e$}
\For{$1$ \KwTo $M$}{
  $V_S \leftarrow$ Sample $S$ synthetic measurements from $\ket{\psi_E}$ using multinomial distribution\;
  $s_s \leftarrow$ \textit{statTest}$(V_S, V_E)$\;
  \If{$s_e \geq s_s$}{
    $c \leftarrow c + 1$\;
  }
}
$p_e \leftarrow c / M$ \tcp*{Compute empirical $p$-value}
\lIfElse{$p_e \ge p_t$}{\Return Pass}{\Return Fail}%
\caption{Monte Carlo version of \StatTest.} %
\label{alg:stats_test_mc}
\end{algorithm}

A commonly used statistical test in quantum software testing literature~\cite{wang2022qusbt,wang2023qucat,UsandizagaMutation2025} is the $\chi^2$ test~\cite{pearson1900x,mcdonald2014handbook}.
To improve robustness and reliability of our study, we also considered the multinomial test~\cite{mcdonald2014handbook}, the G-test (a log-likelihood ratio test), as it is often considered more reliable than the $\chi^2$ test in cases with a small number of observations~\cite{mcdonald2014handbook}, and a Monte Carlo variant of the $\chi^2$ test, Multinomial test, and G-test.  The quantum test outlined in \Cref{alg:stats_test} implements the $\chi^2$ test, G-test, and Multinomial test; and \Cref{alg:stats_test_mc} the Monte Carlo variants of the three statistical tests.  Both start by creating an empty quantum testing circuit with $n$-qubit register and then append the program's input state ($W$) the the program under test ($U$) (lines 1 and 2).  Measurements are then added to all qubits and the testing circuit is executed $S$ times (lines 3 and 4).  Once the testing circuit is finished, measurements are collected and compared against the expected ones (lines 6 to 8 in Algorithm~\ref{alg:stats_test} and lines 6 to 17 in Algorithm~\ref{alg:stats_test_mc}, respectively).

The number of shots $S$ required for these tests depends on the proximity of the states $\ket{\psi_A}$ and $\ket{\psi_E}$. The closer the states are, the harder they are to distinguish, and thus a larger $S$ is needed.
Question is, \emph{what is the minimum number of shots $S$ for a statistical test to be effective?} This is a nontrivial question~\cite{kroonenberg2018tale}. To ensure reliable results, Cochran's rule~\cite[p. 420]{cochran1954some} suggests that each bitstring should have at least $5$ expected observations. Additionally, Pearson's~\cite{pearson1900x} original recommendation requires a minimum of $13$ total observations.

\subsubsection{\swapTest}\label{sec:methodology_swap_test}
Introduced by~\cite{barenco1997,buhrman2001}, measures the similarity between the actual state $\ket{\psi_A}$ and the expected state $\ket{\psi_E}$ by estimating the square of their inner product $\left|\braket{\psi_A | \psi_E}\right|^2$. 
This test is implemented using a quantum circuit as shown in \Cref{fig:swap_test_circuit}, that prepares the two states and applies $n$ controlled-SWAP gates on them (one for each pair of qubits), controlled by an auxiliary qubit $q_a$ initialized to $\ket{0}$. These controlled-SWAP gates are sandwiched between two Hadamard gates applied to the $q_a$.

If the states are orthogonal, the inner product $\left|\braket{\psi_A  | \psi_E}\right|^2 = 0$; if they are identical, $\left|\braket{\psi_A  | \psi_E}\right|^2 = 1$. Thus, as shown in~\cite[p.~167902-2]{buhrman2001}, if the two states are identical, the auxiliary qubit ($q_a$) will always be measured as $0$ (i.e., with probability $P = 1$); if the states are orthogonal, the probability of measuring $q_a$ as $0$ drops to $0.5$: 
\begin{equation}\label{eq:swap_p}
   P(M_{q_a} = 0) =  \frac{1}{2}+\frac{1}{2} \left|\braket{\psi_A  | \psi_E}\right|^2,
\end{equation}
where $M_{q_a}$ denotes measurement on qubit $q_a$.

In typical applications, the \swapTest runs multiple times to estimate the similarity between the states (computed by aggregating the values from the measurements)~\cite{barenco1997,buhrman2001}. However, for the purposes of unit testing, we do not need to compute an estimate of the inner product. Instead, if any measurement returns $1$, we conclude that the states differ and the test case has failed.

The implementation of the \swapTest is provided in \Cref{alg:swap_test}. In practice it is often more efficient to perform $S$ measurements upfront and search the resulting vector for a non-zero value, although one can perform one measurement at a time and stop if a value of $1$ is measured. We adopt the former approach.

The \swapTest is designed to never produce a false positive, as per \Cref{eq:swap_p}: $q_a$ will always be $0$ if the states are equal. However, note that the test can produce a false negative if the value of $S$ is small. As shown in \Cref{eq:swap_p}, the closer the states are, the lower the probability of outcome of $q_a = 1$. In our empirical evaluation (see \Cref{sec:case_study}) we observed that the \swapTest does not produce false positives but can yield false negatives.

\begin{figure}[t]
  \centering
  \begin{tikzpicture}
    \begin{yquant}
      qubit {$\ket{q_a} = \ket{0}$}  a;     %
      qubit {$\ket{\psi_A}$}         qA;    %
      qubit {$\ket{\psi_E}$}         qE;    %
      h a;
      swap (qA, qE) | a;
      h a;
      measure a;
    \end{yquant}
  \end{tikzpicture}
  \caption{Quantum testing circuit for the \swapTest.}
  \label{fig:swap_test_circuit}
\end{figure}
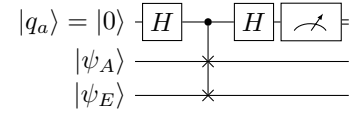
\begin{algorithm}[tb]
\SetKwComment{Header}{}{}
\SetAlgoLined
\SetKwInOut{Parameter}{Parameter}
\Parameter{$W$ as the input state}
\Parameter{$U$ as the software under test}
\Parameter{$\ket{\psi_E}$ as the expected output state}
\Parameter{$S$ as the number of shots}
\SetKwInOut{Input}{input}
\KwResult{Test result: Pass ($U$ behaves as expected) or Fail ($U$ does not behave as expected)}
\Header{\highlightline{yellow!30}{\textbf{Arrange}}}
Create an empty quantum circuit with a $n$-qubit register\;
Append $W$ then $U$ to the circuit\;
Append $\ket{\psi_E}$ to the circuit\;
Append Hadamard and SWAP gates to the circuit as in \Cref{fig:swap_test_circuit}\;
\Header{\highlightline{blue!30}{\textbf{Act}}}
Run the circuit $S$ times\;
$V \leftarrow$ Collect measurement outcomes\;
\Header{\highlightline{red!30}{\textbf{Assert}}}
\lIfElse{$V$ contains only zeros}{\Return Pass}{\Return Fail}
\caption{\swapTest.} %
\label{alg:swap_test}
\end{algorithm}

\subsubsection{\statevectorTest}\label{sec:methodology_quite_statevector_test}

On a classical computer, a simulated (yet accurate) state vector $\ket{\psi_A}$ of a quantum circuit can be computed by sequentially multiplying the unitary matrices that represent its gates\footnote{For example, this can be done using \texttt{Statevector.from\_instruction\\(circuit)} in the Qiskit framework~\cite{javadi2024quantum}.}. This provides direct access to every amplitude in the quantum state, allowing one to compare the simulated state $\ket{\psi_A}$ with an expected state $\ket{\psi_E}$.\footnote{%
Comparing expected and actual state vectors to assess the correctness of a program in a simulator is a common technique among practitioners. Although we could not find a formal academic reference, this approach frequently appears on platforms like StackOverflow, where users mention comparing state vectors to verify the correctness of a program. Here is a typical example: ``For a given unitary, I want to know whether this unitary gate is correctly evolved in the circuit. In the simulator, I can use `statevector' to get the state vector to check the correctness of the evolution.''~\cite{wang2023stackexchange}

From a theoretical perspective, we conjecture that this method is rooted in the concept of \textit{trace distance} in quantum information theory, defined as 
$\frac{1}{2} \Tr|\rho - \sigma|,$
where one compares the density matrices of two states $\rho$ and $\sigma$ to assess their similarity (see~\cite[Sec. 9.2.1]{nielsen_chuang_2010} for details).
Since a density matrix $\rho$ can be computed from a pure state vector $\ket{\psi}$ using $\rho = \ket{\psi} \bra{\psi}$, this may have been the origin of the comparison.%
}

Since the simulation yields the complete quantum state, one can verify the circuit's correctness at a granular level, ensuring not only correct output probabilities but also accurate coherence and phase relationships.
However, this approach becomes computationally infeasible for moderate numbers of qubits, as the state vector grows exponentially with size $2^n$.

Despite this limitation, a \statevectorTest remains a valuable test. Many quantum programs under test involve circuits small enough that the classical simulation is still practical. In such cases, this test is highly reliable, avoiding both false negatives and false positives (within numerical tolerance)\footnote{In this paper, we used a tighter tolerance value of $1 \times 10^{-10}$.  Note this value is lower than the default value of $1 \times 10^{-8}$ defined in numpy's \texttt{np.allclose} function
and in Qiskit's \texttt{equiv} function, because the comparison of identical state vectors in our context requires high precision.}.

\Cref{alg:quite_state_vector} outlines the \statevectorTest. This test offers high accuracy and low run-time, as it requires only a single simulation rather than multiple executions or shots, though it is limited to small quantum registers, as previously mentioned. The algorithm constructs a test circuit by initializing it with $W$, then appending the program under test $U$ (as shown in \Cref{fig:stats_test_circuit}, excluding the measurements). Then it computes the resulting state vector $\ket{\psi_A}$ and compares it with the state vector of $\ket{\psi_E}$. If the vectors match (within tolerance), the test passes, indicating that $U$ behaves as expected.
Notably, in our empirical evaluation (\Cref{sec:case_study}), the \statevectorTest successfully detected all faulty states in the faulty circuits considered.

\begin{algorithm}[t]
\SetKwComment{Header}{}{}
\SetAlgoLined
\SetKwInOut{Parameter}{Parameter}
\Parameter{$W$ as the input state}
\Parameter{$U$ as the software under test}
\Parameter{$\ket{\psi_E}$ as the expected output state}
\SetKwInOut{Input}{input}
\KwResult{Test result: Pass ($U$ behaves as expected) or Fail ($U$ does not behave as expected)}
\Header{\highlightline{yellow!30}{\textbf{Arrange}}}
Create an empty quantum circuit with a $n$-qubit register\;
Append $W$ then $U$ to the circuit\;
\Header{\highlightline{blue!30}{\textbf{Act}}}
$\ket{\psi_A} \leftarrow$ Compute the final state vector of the circuit\;
\Header{\highlightline{red!30}{\textbf{Assert}}}
\lIfElse{$\ket{\psi_A}$ $\approx$ $\ket{\psi_E}$}{\Return Pass}{\Return Fail}%
\caption{\statevectorTest.} %
\label{alg:quite_state_vector}
\end{algorithm}

\subsubsection{\inverseTest}\label{sec:methodology_quite_measurement_test}

Although statistical tests (\Cref{sec:methodology_stats_tests}) only require $n$ qubits to construct the circuit, they involve error-prone test cases.  The \swapTest (\Cref{sec:methodology_swap_test}) effectively eliminates false positives but requires expanding the quantum register from $n$ to $2n + 1$ qubits.  The \statevectorTest (\Cref{sec:methodology_quite_statevector_test}) produces perfect results but lacked scalability.  This raises the question: \emph{can we design a quantum-centric unit test that only uses $n$ qubits, avoids false positives, and scales better than the \statevectorTest?}
The \inverseTest, described below, is our attempt to answer this question.

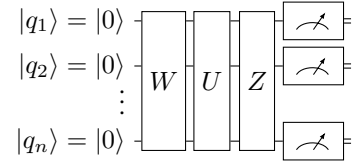
\begin{figure}[tb]
    \centering
    \begin{tikzpicture}
      \begin{yquant}
        qubit {$\ket{q_1} = \ket{0}$} q;
        qubit {$\ket{q_2} = \ket{0}$} q[+1];
        qubit {$\rvdots$} q[+1]; discard q[2];
        qubit {$\ket{q_n} = \ket{0}$} q[+1];
        box {$W$} (q);
        box {$U$} (q);
        box {$Z$} (q);
        measure q[0,1,3];
      \end{yquant}
    \end{tikzpicture}
    \caption{Quantum testing circuit for the \inverseTest.}
    \label{fig:quite_test_strategy}
\end{figure}
\begin{algorithm}[t]
\SetKwComment{Header}{}{}
\SetAlgoLined
\SetKwInOut{Parameter}{Parameter}
\Parameter{$W$ as the input state}
\Parameter{$U$ as the software under test}
\Parameter{$\ket{\psi_E}$ as the expected output state}
\Parameter{$S$ as the number of shots}
\SetKwInOut{Input}{input}
\KwResult{Test result: Pass ($U$ behaves as expected) or Fail ($U$ does not behave as expected)}
\Header{\highlightline{yellow!30}{\textbf{Arrange}}}
Create an empty quantum circuit with a $n$-qubit register\;
Append $W$ then $U$ to the circuit\;
Append the conjugate transpose (inverse) of $\ket{\psi_E}$ as $Z$ to the circuit\;
Add final measurements to all qubits in the circuit\;
\Header{/* At this point, the circuit looks like the one depicted in \Cref{fig:quite_test_strategy} */}
\Header{\highlightline{blue!30}{\textbf{Act}}}
Transpile the circuit for a backend (either a simulator or an actual device)\;
Run the transpile circuit $S$ times\;
$V \leftarrow$ Collect measurement outcomes\;
\Header{\highlightline{red!30}{\textbf{Assert}}}
\lIfElse{$V$ contains only zero-bitstrings, i.e., $(00\cdots0)$}{\Return Pass}{\Return Fail}%
\caption{\inverseTest.} %
\label{alg:quite_measurement}
\end{algorithm}

The core idea is to revert\footnote{We are inspired by prior works that explore the reversibility of circuits, e.g.,~\cite{1317002,6231097,7977062,1197682,syamala2012,patel2021qraft,MondalReversible2022}, conducted in different contexts and complementary to our approach.} the output state $\ket{\psi_A}$ back to $\statezero$, effectively reducing the circuit output to a single deterministic value, namely, the zero-bitstring. This is achieved by constructing a unitary operator $Z$ such that:
$$\ket{\psi_R} = \statezero \longleftarrow Z \ket{\psi_A}.$$

Obviously, if $U$ is implemented correctly, then $\statezero  \longleftarrow  W^\dagger U^\dagger \ket{\psi_A}$, where $\dagger$ superscript denotes the conjugate transpose. While $W^{\dagger}$ can be computed from $W$ using frameworks like Cirq~\cite{cirq_developers_2024_11398048}, PennyLane~\cite{bergholm2018pennylane}, or Qiskit~\cite{javadi2024quantum}, computing conjugate transpose of $U$ directly would not reveal defects in $U$. 
Instead, $Z$ is constructed using the conjugate transpose of the expected state $\ket{\psi_E}$, ensuring that $\ket{\psi_R} = \statezero$, only if $U$ is correct. This works because quantum operations are reversible~\cite{nielsen_chuang_2010}: applying the complex conjugate of a unitary operation to its result restores the original state. 
After constructing the testing circuit, we can perform repeated measurements\footnote{If necessary, the measurement of $\statezero$ can be reduced to $\ket{0}$ using one additional gate and an auxiliary qubit. After appending the $Z$ operator in \Cref{fig:quite_test_strategy}, add a multi-controlled gate with zero-controls on $q_1, q_2, \ldots, q_n$. This gate flips the auxiliary qubit only if all $n$ control qubits are in the $\ket{0}$ state. As a result, the auxiliary qubit measures $\ket{1}$ if the test passes, and $\ket{0}$ if it fails.} to assess its result.  Unlike the \statevectorTest (\Cref{sec:methodology_quite_statevector_test}), this test uses actual circuit measurements, making it more scalable for complex programs.

\Cref{alg:quite_measurement} outlines the \inverseTest. It constructs a test circuit, initializes it with $W$ and appends the program under test $U$, constructing $\ket{\psi_A}$. Then, it appends conjugate transpose (inverse) of $\ket{\psi_E}$. The circuit is then transpiled for a given backend (simulator or hardware), run for multiple shots, and the output measurements $V$ are collected. If all observed bitstrings are zero, the test \emph{passes}.

As for the \StatTests and the \swapTest, the question is, \emph{how many shots $S$ are needed to confidently conclude that non-zero string is never measured?} This depends on the closeness of the faulty state to the correct state. The nearer the two states are, the more shots are required. 
Specifically, the number of required shots grows exponentially as the faulty state approaches the correct state\footnote{As an analogy, consider two coins: one fair ($p = 0.5$ for heads), and one slightly biased ($p = 0.5 + 10^{-10}$). Distinguishing them reliably would require a very large number of flips.}.
We formally analyze this exponential growth %
with the help of the Quantum Chernoff Bound, denoted $\xi_{\text{QCB}}$ (see~\cite{audenaert2007discriminating,nussbaum2009chernoff} for further details). 
As per \cite[Eqs.~1 and~2]{audenaert2007discriminating},
\begin{equation}\label{eq:quantum_chernoff_bound}
\begin{split}
    P_e &\sim \exp(-N \xi_{\text{QCB}}), \quad \text{where}\\
    \xi_{\text{QCB}} &= \lim_{N \to \infty} - \frac{\ln P_e}{N}  \\
    &= - \ln \min_{0 \leq s \leq 1} \Tr \left( \rho^s \sigma^{1-s} \right) 
\end{split}
\end{equation}
and $P_e$ is the error probability, $N$ is the number of shots, $\Tr$ denotes the matrix trace, $\rho$ is the density matrix representing a passing test case (the zero-state), and $\sigma$ is the density matrix for the buggy state that differs from the zero-state.

From \Cref{eq:quantum_chernoff_bound}, given $\rho$ and $\sigma$, we can compute the required number of shots $N$ for a specified error probability $P_e$ as:
\begin{equation}\label{eq:estimate_of_n}
    N \sim \frac{\ln{\left(P_e\right)}}{\ln \min_{0 \leq s \leq 1}  \left[ \Tr \left( \rho^s \sigma^{1-s} \right) \right]}.
\end{equation}

\paragraph{Numerator Analysis of \Cref{eq:estimate_of_n}}
Technically, $P_e \in (0, 1)$, but in practice a tester would usually use $0 < P_e \le 0.05$. As $P_e$ decreases, the absolute value of the numerator $\ln(P_e)$ increases.

\paragraph{Denominator Analysis of \Cref{eq:estimate_of_n}}
The density matrix $\rho$ for a passing test case (zero-state) is given by:
$$
\rho = \ket{0}^{\otimes n} \bra{0}^{\otimes n} = \begin{bmatrix}
1 & 0 &  \cdots & 0 \\
0 & 0 &  \cdots & 0 \\
\vdots &  \vdots & \ddots & \vdots \\
0 & 0 & \cdots & 0
\end{bmatrix}.
$$
Matrix $\rho$ contains $2^n \times 2^n$ elements, with only one non-zero element ($\rho_{11}$).
Since $\rho^s = \rho$ for any $s$, \Cref{eq:estimate_of_n} simplifies to:
\begin{equation}\label{eq:estimate_of_n_rho_zero}
    N \sim \frac{\ln{\left(P_e\right)}}{\ln \min_{0 \leq s \leq 1}  \left[ \Tr \left( \rho \sigma^{1-s} \right) \right]}.
\end{equation}

However, the structure of $\sigma$ can vary significantly, making the computation of $N$ challenging for arbitrary cases. Below, we analyze specific cases of $\sigma$ to build intuition about the behavior of the denominator.

\bigskip\noindent\emph{(a) Diagonal Pure States}

\smallskip\emph{Case 1: $\sigma$ is diagonal with $\sigma_{11} = 0$.}
Consider a pure state where the density matrix $\sigma$ is diagonal, i.e., $\sigma_{ij} = 0$ for all $i \neq j$, and $\sigma_{11} = 0$. The remaining diagonal elements sum to 1:
$$
\sigma = 
\begin{bmatrix}
0 & 0 & \cdots & 0 \\
0 & \sigma_{22} & \cdots & 0 \\
\vdots & \vdots & \ddots & \vdots \\
0 & 0 & \cdots & \sigma_{mm}
\end{bmatrix}, \quad \text{with } \sum_{i=2}^m \sigma_{ii} = 1,
$$
where $m=2^n$. In this case:
$$
\Tr(\rho \sigma^{1-s}) = 1.
$$
Substituting into \Cref{eq:quantum_chernoff_bound} yields:
\begin{equation}
    P_e \sim \exp \left( -N \cdot 0 \right) = 1.
\end{equation}
This indicates that for trivial cases where $\rho$ and $\sigma$ are orthogonal, a single shot suffices to differentiate them.

\begin{figure}[tb]
    \centering
    \includegraphics[width=0.8\columnwidth]{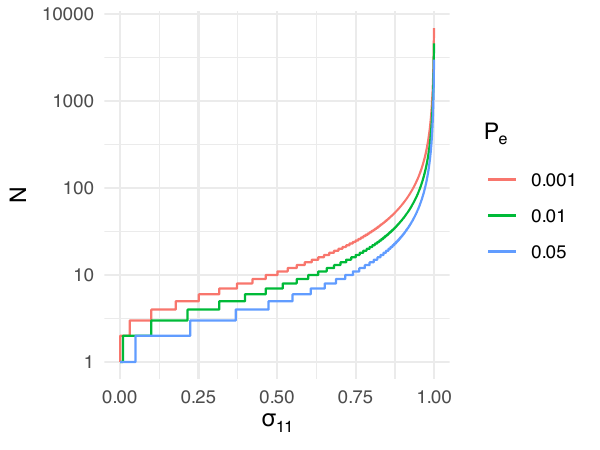}
    \vspace{-1em}\caption{The value of $N$ from \Cref{eq:estimate_of_n_diag_case_ceiling} for $0.001 \leq \sigma_{11} \leq 0.999$ and $P_e \in \{0.001, 0.01, 0.05\}$.} 
    \label{fig:estimate_of_n_diag_case_line}
\end{figure}

\smallskip\emph{Case 2: $\sigma$ is diagonal with $0 < \sigma_{11} < 1$.}
Now consider $\sigma$ where $0 < \sigma_{11} < 1$ and some other diagonal elements are non-zero, such that $\Tr(\sigma) = 1$. In this case:
$$
\Tr(\rho \sigma^{1-s}) = \sigma_{11}^{1-s}.
$$
To minimize $f(\sigma_{11}, s) = \ln (\sigma_{11}^{1-s})$, we compute its derivative and solve it for zero:
$$
\diffp{f(\sigma_{11},s)}{s} = -\ln(\sigma_{11})^{1 - s} \ln[\ln(\sigma_{11})] = 0.
$$
Since $\sigma_{11} \in (0,1)$, there are no values of $s$ that satisfy this equality, i.e., we do not have a critical point when  $s \in [0,1]$. Thus, the minimum must occur at one of the boundaries of the domain for $s$:
\begin{itemize}[leftmargin=*]
    \item At $s = 0$: $f(\sigma_{11},0) = \ln(\sigma_{11})^1 = \ln(\sigma_{11})$,
    \item At $s = 1$: $f(\sigma_{11},1) = \ln(\sigma_{11})^0 = 1$.
\end{itemize}
As $\sigma_{11} \in (0,1)$, $\ln(\sigma_{11}) <0$ and the minimum is reached when $s=0$. Substituting $\ln(\sigma_{11})$ into \Cref{eq:estimate_of_n_rho_zero} yields:
\begin{equation}\label{eq:estimate_of_n_diag_case}
    N \sim \frac{\ln{\left(P_e\right)}}{\ln \min_{0 \le s \le 1}  \left[ \Tr ( \rho \sigma^{1-s}\right)]} = \frac{\ln{\left(P_e\right)}}{\ln (\sigma_{11})}.
\end{equation}
In practice, we round up $N$ to the nearest integer away from zero, ensuring there is at least one shot:
\begin{equation}\label{eq:estimate_of_n_diag_case_ceiling}
    N \sim \max \left\{ \left\lceil \frac{\ln{\left(P_e\right)}}{\ln(\sigma_{11})} \right\rceil, 1 \right\},
\end{equation}
where $\lceil \cdot \rceil$ denotes the ceiling function. \Cref{fig:estimate_of_n_diag_case_line} shows how $N$ in \Cref{eq:estimate_of_n_diag_case_ceiling} changes with $\sigma_{11}$ and $P_e$. As $\sigma_{11} \to 1$, $N$ increases exponentially. Conversely, as $\sigma_{11}$ decreases, $N$ quickly approaches 1. The value of $P_e$ also contributes to $N$, which increases as $P_e$ decreases. However, $\sigma_{11}$ has a stronger influence on $N$ compared to $P_e$.

\bigskip\noindent\emph{(b) Non-Diagonal Pure States and Mixed States}\label{sec:n_diag_states}

\smallskip When non-diagonal elements are present in $\sigma$, symbolic analysis becomes challenging and often requires numerical methods. The complexity increases further with mixed states\footnote{Mixed state $\sigma$ is generally harder to distinguish from $\rho$ because they may share partial support in the same subspace, leading to a higher shot count.}. However, the same principle holds: $N \to \infty$ as $\sigma \to \rho$. Intuitively, as the two states become more similar, more measurements are needed to differentiate them. However, if $\sigma$ is very close but not identical to $\rho$, this approximate similarity may be sufficient for certain practical use cases where perfect discrimination is unnecessary. This is analogous to scientific computing, where computations often involve approximations and numerical errors due to factors such as finite-precision arithmetic. In such contexts, it is generally acceptable to tolerate a small deviation from the exact solution, provided that the error remains within a pre-specified tolerance level. Similarly, the near-equivalence of $\sigma$ and $\rho$ can still yield meaningful and reliable results when the application does not require absolute precision.

To develop an intuition for the shot count $S$, which is defined using the same principles as \Cref{eq:estimate_of_n_diag_case_ceiling}, we express it as  
\begin{equation}\label{eq:estimate_of_s}
    S \sim \max \left\{ \left\lceil \frac{\ln{( P_e)}}{\ln \min_{0 \leq s \leq 1}  \left[ \Tr \left( \rho \sigma^{1-s} \right) \right]} \right\rceil, 1 \right\}.
\end{equation}  
$S$ varies for an arbitrary state $\sigma$; thus, it is useful to analyze the behavior of $S$ on a range of empirical circuits (which we perform in \Cref{sec:case_study}). This is needed because $S$ gives us an approximation rather than the exact shot count.

\smallskip In summary, our analysis shows that a single shot suffices to distinguish orthogonal states, while distinguishing nearly identical states may require $> 10^{7}$ shots. We confirm this behavior empirically in \Cref{sec:case_study}.

\subsection{Analysis of complexity}\label{sec:complexity_analysis}

\begin{table*}[tb]
\centering
\caption{Comparison of the non-quantum-centric and quantum-centric unit tests described in \Cref{sec:methodology_qut}.\\
{\small Functions $D(x)$ and $G(x)$ denote the circuit depth and gate count, respectively, for implementing unitary operators or states $x$. Useful relations include: $D(\ket{\psi_A}) = D(UW)$, $D(\ket{\psi_E}) = D(\ket{\psi_E}^\dagger) = D(Z) = D(Z^\dagger)$, $G(\ket{\psi_A}) = G(U) + G(W)$, and $G(\ket{\psi_E}) = G(\ket{\psi_E}^\dagger) = G(Z) = G(Z^\dagger)$. The approximations shown, denoted by $\approx$, hold only if the circuits for $\ket{\psi_A}$ and $\ket{\psi_E}$ are nearly identical  in terms of depth and gate count.}}\label{tbl:theoretical_comparison_of_tests}
\vspace{-1em}\resizebox{\textwidth}{!}{%
\begin{tabular}{@{\extracolsep{\fill}}p{5cm}llll}
\toprule
                                   & \textbf{\StatTests} & \textbf{\swapTest} & \textbf{\statevectorTest} & \textbf{\inverseTest} \\
\midrule
Circuit width                                              & $n$ & $2n+1$ & $n$ & $n$ \\
Circuit depth                                              & $D(\ket{\psi_a})$ & $\max(D(\ket{\psi_a}), D(\ket{\psi_E})) + n + 2 \approx D(\ket{\psi_a})+n$ & $D(\ket{\psi_a})$ & $D(\ket{\psi_R}) \approx 2 D(\ket{\psi_a})$ \\
Circuit number of gates                                         & $G(\ket{\psi_A})$ & $G(\ket{\psi_A}) + G(\ket{\psi_E}) + n + 2 \approx 2(G(\ket{\psi_A})) + n$ & $G(\ket{\psi_A})$ & $G(\ket{\psi_A}) + G(\ket{\psi_E}) \approx 2(G(\ket{\psi_A}))$ \\
Shot count ($S$)                                          & $\ge 13$ & $\geq 1$ & $0$ & $\geq 1$ \\
Can produce false-positives?                                                 & Yes & No & No & No \\
Can produce false-negatives?                                                 & Yes & Yes & No & Yes \\
Classical computer prep. time when $\ket{\psi_A}$ is a circuit and $\ket{\psi_E}$ is not & $O(G(\ket{\psi_A}) + S(\ket{\psi_E}))$~$^{*}$ & $O(G(\ket{\psi_A}) + C(\ket{\psi_E}) +n)$~$^{\mathsection}$ & $O(2^n)$ & $O(G(\ket{\psi_A})+C(\ket{\psi_E}))$~$^{\mathsection}$ \\
Classical computer prep. time when $\ket{\psi_A}$ and $\ket{\psi_E}$ are circuits     & $O(G(\ket{\psi_A}) + 2^n)$~$^{*}$ & $O(G(\ket{\psi_A}) + G(\ket{\psi_E})+n)$ & $O(2^n)$ & $O(G(\ket{\psi_A}) + G(\ket{\psi_E}^\dagger))$ \\
Classical computer simulator time                                                    & $O(2^n)$ & $O(2^{2n+1})$ & $O(2^n)$ & $O(2^n)$ \\
Quantum computer time       & $O(D(\ket{\psi_a}))$ & $O(\max(D(\ket{\psi_a}), D(\ket{\psi_E}))) \approx O(D(\ket{\psi_a}))$ & $0$ & $O(D(\ket{\psi_R})) \approx O(D(\ket{\psi_a}))$~$^{\ddagger}$ \\
\bottomrule
\multicolumn{5}{p{22cm}}{$^{*}${If the  state vector for $\ket{\psi_E}$ is dense, sampling the vector $V_E$ (see \Cref{alg:stats_test}) requires $O(2^n)$ operations. Thus, the computational complexity for sampling $S(\ket{\psi_E})$ is $O(2^n)$.} However, if the state vector is sparse, the complexity drops to $S(\ket{\psi_E}) = O(r)$, where $r$ is the number of nonzero elements in $\ket{\psi_E}$ state vector. This significantly reduces preparation time.} \\
\multicolumn{5}{p{22cm}}{$^{\ddagger}$ {Quantum computer time for the \inverseTest will be twice that of the Statistical and \swapTest, assuming the circuits for $\ket{\psi_A}$ and $\ket{\psi_E}$ are similar.}} \\
\multicolumn{5}{p{22cm}}{$^{\mathsection}$ If $\ket{\psi_E})$ is a dense state vector, converting it to a circuit has computational complexity $C(\ket{\psi_E}) = O(2^n)$. For a sparse state vector, the complexity reduces to $C(\ket{\psi_E}) = O(nr)$, where $r$ is the number of nonzero elements in $\ket{\psi_E}$ statevector.}\\
\end{tabular}
}
\end{table*}

\Cref{tbl:theoretical_comparison_of_tests} compares the complexity of the four quantum-centric unit tests. We focus on a common practical case in which the \bug is detected, making the circuits for $\ket{\psi_A}$ and $\ket{\psi_E}$ nearly identical in terms of depth and gate count.

\subsubsection{Circuit Width}
Among all methods, the \swapTest requires the widest circuit, using $2n+1$ qubits, i.e., roughly twice as many as the others. Since execution costs grow exponentially with the qubit count, this makes the \swapTest the most expensive to simulate and potentially more difficult to execute on early fault-tolerant quantum devices, where the number of available qubits will be limited.

\subsubsection{Circuit Depth and Gate Count}
The \inverseTest requires the deepest circuit (approximately twice the depth of the other methods) leading to longer execution times on a quantum device. Both the \swapTest and \inverseTest have roughly twice the gate count compared to the \StatTests and \statevectorTest.

\subsubsection{Shot Count}
\StatTests require multiple shots. Although the \swapTest and \inverseTest also require repeated execution, they might potentially require fewer shots. In contrast, the \statevectorTest is deterministic and does not require any quantum execution or simulation, i.e., it can be computed purely through classical matrix multiplications.

\subsubsection{False Positives/Negatives}
The \statevectorTest is the most reliable, producing neither false positives nor false negatives. \StatTests are the least reliable, susceptible to both types of errors. The \swapTest and \inverseTest lie in between~---~they avoid false positives but can still yield false negatives.

\subsubsection{Classical Compute Time (Preparation)}
\paragraph{When $\ket{\psi_A}$ is a circuit and $\ket{\psi_E}$ is a state vector} If the state vector $\ket{\psi_E}$ is dense, all methods incur exponential cost due to the size of the vector ($2^n$). For the \statTests, this corresponds to sampling from a distribution over $2^n$ elements. In the \statevectorTest, $\ket{\psi_E}$ is directly available, but it requires handling $2^n \times 2^n$ matrix multiplications representing $\ket{\psi_A}$. For the \swapTest and \inverseTest, converting the classical state vector $\ket{\psi_E}$ into a quantum circuit requires circuit synthesis, which also incurs exponential cost~\cite[Sec.4.5.4]{nielsen_chuang_2010}; see~\cite{miranskyy2024comparing} for an overview of conversion techniques.

However, if $\ket{\psi_E}$ is sparse with only $r$ non-zero elements ($2^n \gg r$), the \statTests requires only $O(r)$ operations to sample from the distribution. The cost of the \statevectorTest remains unchanged, as the representation of $\ket{\psi_A}$ is unaffected. For the \swapTest and the \inverseTest, the cost of synthesizing a circuit for $\ket{\psi_E}$ drops to $O(nr)$; see~\cite{miranskyy2024comparing} for a review of conversion methods.  In this case, the \statTests has the lowest computational complexity, followed by the \inverseTest, the \swapTest, and finally the \statevectorTest.

\paragraph{When both $\ket{\psi_A}$ and $\ket{\psi_E}$ are circuits}
This scenario can occur during post-optimization validation. For instance, suppose we have an existing circuit, $\ket{\psi_E}$. After optimizing it, we obtain a new version, $\ket{\psi_A}$, and we want to ensure that the optimized circuit still produces the correct state.

Classical preparation time for the \statTests and the \statevectorTest remains unchanged. For the \swapTest and \inverseTest, prep time drops from exponential to linear, since the circuit for $\ket{\psi_E}^\dagger$ can be obtained by inverting the original circuit (which is linear in gate count).

In this case, the \inverseTest becomes the most efficient, followed by the \swapTest, particularly on fault-tolerant quantum hardware. While simulation costs still grow exponentially, the preparation and quantum execution times become manageable, making it suitable for larger circuits.

\subsubsection{Simulator and Quantum Hardware Execution}
The \swapTest remains the most resource-intensive to simulate due to its circuit width. In contrast, the \StatTests, the \statevectorTest, and the \inverseTest operate on $n$-qubit registers, resulting in lower simulation costs. On quantum hardware, \StatTests are the fastest, followed by the \swapTest. The \inverseTest requires circuits that are twice as deep. However, for the all-circuit case, discussed above, it is the most practical option for fault-tolerant devices, provided that $\ket{\psi_A}$ and $\ket{\psi_E}$ can be prepared efficiently on a classical computer and testers aim to avoid false positives.

\subsubsection{Summary}
When $n$ is small, the \statevectorTest executed on a classical computer yields the most reliable results, with exact answer and no false positives or negatives. If large $n$ cannot be avoided and the test has to be executed on a fault-tolerant quantum computer, the \inverseTest is the best option (for a dense state vector $\ket{\psi_E}$). If $n$ is large and $\ket{\psi_E}$ is a sparse state vector, then \StatTests are the cheapest  option (although the programmer will suffer from false positives and negatives). The \inverseTest and the \swapTest are more expensive alternatives in this case. However, while they require more classical computation time (and in the case of the \swapTest, twice as many qubits) they eliminate concerns about false positives.

\section{Empirical study}\label{sec:case_study}

This paper aims to %
evaluate several quantum unit tests on the 
following research questions (RQs):
\begin{enumerate}[label=\textbf{RQ\arabic*:}, leftmargin=*]
  \item Which testing method performs best at reliably distinguishing between quantum states that are different?

  \item Which testing method requires the fewest shots to detect true positives while keeping false negatives low?

\end{enumerate}

\subsection{Defects}\label{sec:evaluation:operators}

In this work, we focus on defects in the program code written by the developer, assuming that the rest of the stack is free of defects.
To simulate defects one might introduce, we applied four mutation operators, one at time, to a quantum circuit, either randomly-generated or real.
Three mutation operators (i.e., QGR or simply Replace, QGD or Remove, and QGI or Add) have been proposed~\cite{QMutPyJournal,Muskit} and we, in this study, propose a new one (i.e., RGI).  Each operator is presented below.
Note that \citet{QMutPyJournal} has also proposed mutation operators QMI and QMD but neither were not included in our study as both operators would only generate equivalent mutants.
The former operator introduces measurements at any point in the circuit, which
would then be removed by our experimental scripts before computing circuit's state
vector (e.g., in the \statevectorTest) as it requires a measurement-free circuit.
In other words, once the all measurements are removed from the original and the
mutated circuits, the circuits are equivalent.
The latter, which deletes measurements from the circuit, would also be pointless.
Suppose there is a circuit with four measurements.  The QMD operator would
randomly select a measurement and delete it.  Our experimental scripts, would then remove
all measurements in the original circuit (i.e., four) and all the remaining measurements in the
mutated circuits (i.e., three).  Thus, both circuits would be equivalent.

\smallskip\textit{Quantum Gate Replacement (QGR)} operator replaces an existing quantum gate call in the circuit (e.g., {\small\texttt{circuit.x(qubit)}}) with each of its syntactically equivalent gates\footnote{A syntactically equivalent quantum gate, as defined by \citet{QMutPyJournal}, is a gate that can replace another in code without causing syntax errors, even if it performs a different quantum operation.} (e.g., {\small\texttt{circuit.h(qubit)}}, {\small\texttt{circuit.s(qubit)}}, etc.), generating a mutant per replacement.

\smallskip\textit{Quantum Gate Deletion (QGD)} mimics an error where a programmer accidentally removes a gate from the circuit.

\smallskip\textit{Quantum Gate Insertion (QGI)} performs the opposite of QGD and inserts an additional call to a syntactically equivalent quantum gate immediately after an existing gate call (e.g., adding {\small\texttt{circuit.y(qubit)}} after {\small\texttt{circuit.x(qubit))}}, creating mutants for each possible equivalent gate.

\smallskip\textit{Rotation Gate Insertion (RGI)} simulates an error where a programmer introduces a rotational gate\footnote{RGate documentation: \url{https://docs.quantum.ibm.com/api/qiskit/1.3/qiskit.circuit.library.RGate}, accessed July 2025.}:
$$\text{RGate}=\exp{\left[-i \frac{\theta}{2}(\cos \phi x + \sin \phi y)\right]}, $$ at an arbitrary position in the circuit, with a small rotation deviation of $\theta = \phi = \frac{\pi}{180}$ (i.e., one-degree angle).

\subsection{Subjects}

We apply mutation operators described in \Cref{sec:evaluation:operators} to two datasets of quantum circuits discussed below.
To conserve computational resources, since memory and computation costs grow exponentially with the number of qubits $n$, we capped $n$ at 5 for both datasets. %

\paragraph{MQT Bench} The Munich Quantum Toolkit Benchmark Library (MQT Bench)\footnote{MQT Bench homepage: \url{https://www.cda.cit.tum.de/mqtbench}, accessed July 2025.}~\cite{quetschlich2023mqtbench}~v1.1.9 is composed of 1,943  quantum programs (implementing various quantum algorithms) of different sizes.  
Restricting our analysis to $n = 5$ qubits yields 85 circuits with a number of qubits spans between 2 and 5 and circuit depth varies between 1 and 403 instructions.  We then mutate each circuit.  Due to time constraints, for each circuit and mutation operator, we only consider 10\% (selected at random) of all possible mutants.  This generates a total of 45,030 mutated circuits.

\paragraph{Randomly-generated circuits} To increase the number and diversity of the quantum circuits used in our empirical study, we randomly generate circuits using Qiskit's \texttt{random\_circuit}\footnote{Qiskit Random Circuits documentation: \url{https://quantum.cloud.ibm.com/docs/en/api/qiskit/circuit_random}, accessed July 2025.} module.  We randomly generate 1,000 \emph{different} circuits for all combinations of number of qubits 1, 2, 3, 4, and 5; and depth 1 and 10.  This results in 10,000 circuits which we then mutate to generate a total of 1,751,850 mutated circuits.

\smallskip Note, \emph{no mutated circuit} in either the MQT Bench or the Randomly-generated circuits dataset \emph{is equivalent to its correspondent original circuit}. Specifically, we removed any pair for which the original and mutated circuits produced identical statevectors (within a numerical tolerance of $10^{-10}$). There were 2,189~$\leftarrow$~47,219~$-$~45,030 equivalent pairs in the MQT Bench and none in the Randomly-generated circuits dataset (as non-equivalent pairs were generated by design). This filtering ensures that all mutants in our evaluation represent semantically distinct behaviors. As a result, we do not have any negative cases in our study.

Also note that while our filtering removes equivalent \emph{original-mutant} pairs, it is possible that \emph{two different mutants} of the same circuit produce identical state vectors. We do not account for such cases in this work, but acknowledge this as an avenue for future investigation.

\subsection{Experimental approach}\label{sec:execAndmeas}

\subsubsection{Setup}\label{sec:setup}

\paragraph{Circuits' inputs} The quantum circuits are initialized, in each quantum unit test, with the default input state, i.e., $\statezero$.

\paragraph{Shot count}\label{sec:shot_count} The maximum number of shots is set to $10^6$ for the MQT Bench and to $10^4$ for the Random Circuits dataset. For individual experiments (i.e., a test on a pair original-mutant), the number of shots is capped at twice the theoretical limit given by \Cref{eq:estimate_of_s} with $P_e = 0.05$. This cap is applied across all tests, although the analysis focuses specifically on the \inverseTest. Since our goal is to identify the top-performing test method rather than to conduct an in-depth analysis of each method, this approach is sufficient. 

\paragraph{Repetitions} To increase reproducibility, each experiment (i.e., a test on a pair original-mutant) is repeated 100 times, each time with a different seed value,
as the sequence of obtained measurements varies with each execution.\footnote{The recommendations in classical software engineering is to run experiments that might have non deterministic results at least 30 repetitions~\cite{Hitchhiker12}.  We chose a higher value to further increase the confidence in our results.} Thus, the total number of experiments is therefore
$1,585,665,000 = (1,751,850+10,000)~\text{pairs} \times 9~\text{tests} \times 100~\text{repetitions}$.

\paragraph{Number of repetitions in the Monte Carlo statistical tests}\label{sec:mc_reps} To conserve computational resources, we set the number of Monte Carlo repetitions $M$ in \Cref{alg:stats_test_mc} to $1000$. Given $P_e = 0.05$, we use a consistent $p$-value threshold of $p_t = 0.05$ for the Monte Carlo tests. The standard error (SE) of the Monte Carlo estimate, assuming a binomial distribution, is then
$$\text{SE} = \sqrt{\frac{p_t(1-p_t)}{M}} = \sqrt{\frac{0.05 \cdot 0.95}{1000} } \approx 0.007. $$

\paragraph{Runtime environment} The experiments were carried out on a high performance computing cluster equipped with Intel Xeon Gold 6148 Skylake and AMD EPYC 9654 Zen 4 CPUs. Each experiment was allocated 16~GB of memory and 2~CPU cores. The total computational cost for the ``production'' run was approximately 9.3~CPU-years. The code was executed using Python~v.3.11.5 with packages qiskit~v.1.4.0, qiskit-aer~v.0.16.1, scipy~v1.16.0, and numpy~v2.3.1.

\subsubsection{Metrics}

\paragraph{True Positive (TP), False Negative (FN), and Recall} For each test, we report the number of true positives, the number of false negatives, and the recall. Since these experiments are designed without negative cases, related metrics such as accuracy and precision are omitted.

\paragraph{Determining a number of shots for a given observation} To determine the number of shots for a given  for the Swap and Inverse tests, we simply identify the first non-zero measurement string in the vector of measured values.

For the statistical test, our aim is to find the minimum number of shots required for the $p$-value to fall below a specified threshold $p_t=0.05$. Typically, $p$-values start high with low shot counts and decrease as the shot count increases, though not always monotonically. A naive approach would increase the shot count from 1 until the $p$-value drops below the threshold, but this is computationally expensive. Instead, we use a two-pass search strategy:
\begin{enumerate*}
    \item Perform a coarse scan to identify a range of shot counts where the $p$-value crosses the threshold;
    \item Use binary search within that range to find the exact minimum shot count.
\end{enumerate*}

\paragraph{Ranking tests}
When comparing the number of shots, we use a dense rank approach. For example, if four test methods require 3, 7, 8, and 7 shots, respectively, their ranks would be 1, 2, 3, and 2.

\subsubsection{Procedure}
For each pair of expected and actual circuits (that is, expected and mutated circuits) we compare their state vectors to evaluate the \statevectorTest. This comparison is done once per pair.

For the remaining tests, we construct the circuits shown in \Cref{fig:stats_test_circuit,fig:swap_test_circuit,fig:quite_test_strategy} and perform the specified number of measurements for each, as outlined in paragraph~\ref{sec:shot_count}. We then post-process the results to determine whether each test passed or failed (following the algorithms described in \Cref{sec:methodology_qut}). This entire process is repeated 100 times to increase robustness, as the observed measurement sequences vary with each run.

\subsection{Results and Answers to RQs}\label{sec:case_study_results}

\begin{figure}
    \centering
    \includegraphics[width=\linewidth]{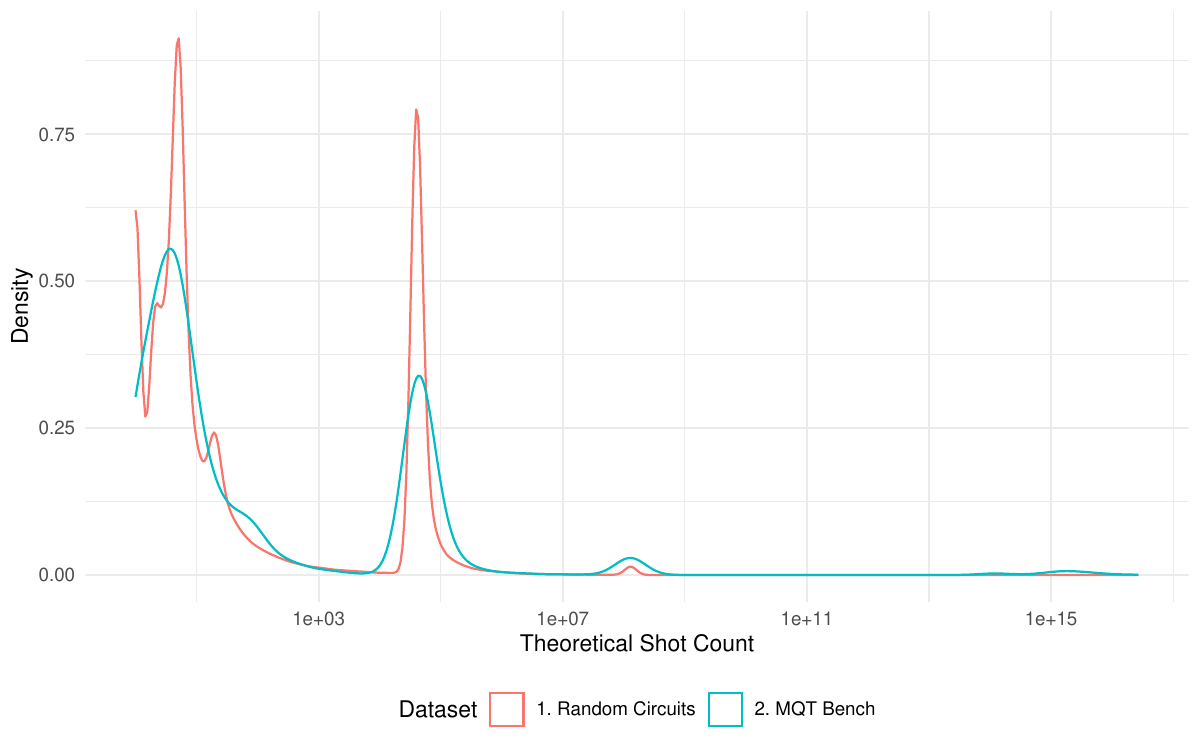}
    \caption{Theoretical shot count $S$ based on \Cref{eq:estimate_of_s} with $P_e=0.05$.}
    \label{fig:dataset_theor_shot_count}
\end{figure}

\smallskip As shown in \Cref{fig:dataset_theor_shot_count}, while many experiments require fewer than $10^3$ shots to reliably distinguish between quantum states that are different, some require up to $10^5$ shots. The distribution also has heavy tails; for example, some cases require more than $10^{15}$ shots to detect a difference. Both datasets show similar distribution profiles, which is expected given the use of comparable mutation operators that produce similar differences between the actual and expected states.

\Cref{fig:ranking_dashboard} shows that, for the MQT Bench dataset, the \textbf{\inverseTest requires the fewest shots (rank-1) in $\approx$69\% of cases}, followed by the \swapTest at $\approx$33\%. The multinomial test ranks third with $\approx$17\%, but results in more false negatives (and is known to be prone to false positives as we saw in \Cref{sec:intro}), making it potentially risky to use. The remaining statistical tests show similar behavior but perform worse overall.

For the Random Circuits dataset in \Cref{fig:ranking_dashboard}, a similar trend is observed. The \textbf{\inverseTest performs best, achieving rank-1 results in $\approx$78\% of cases}, followed by the \swapTest at $\approx$40\%. The statistical tests perform similarly to each other, with rank-1 results below 1\%.

\begin{figure}[tb]
    \centering
    \includegraphics[width=\linewidth]{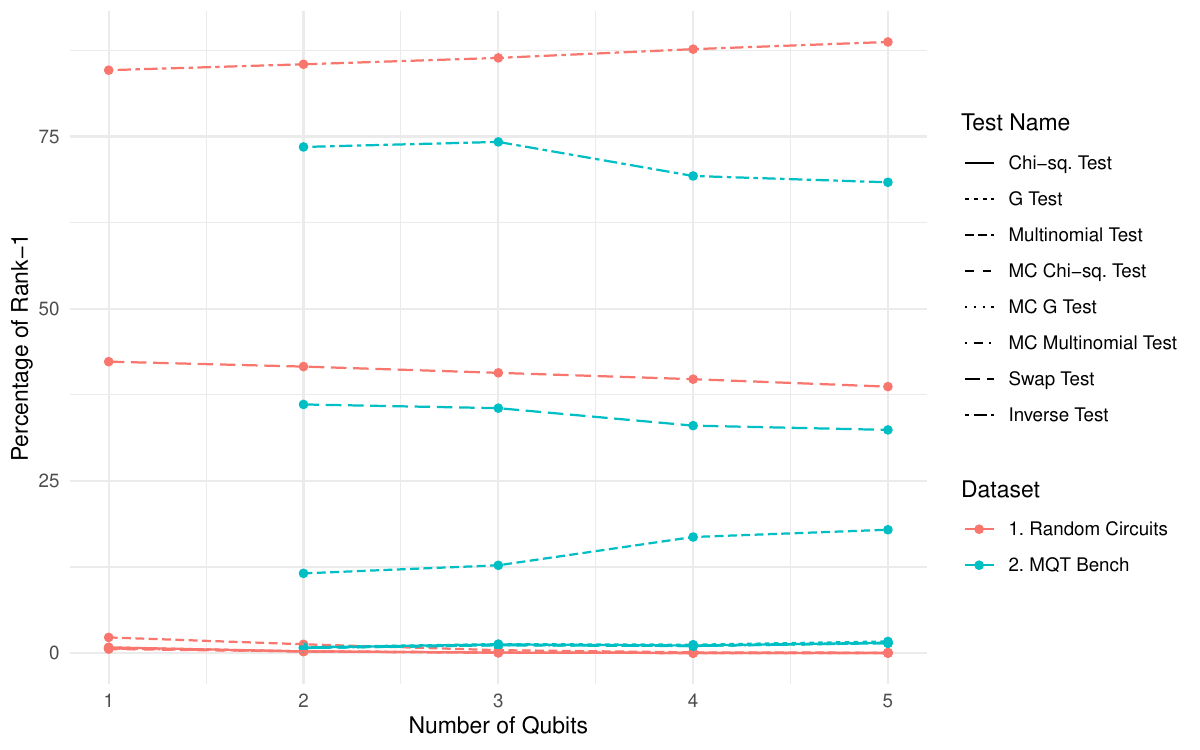}
    \caption{Variation in the percentage of rank-1 cases with the number of qubits.}
    \label{fig:rank_1_per_num_qubits}
\end{figure}

\begin{figure*}[tb]
  \centering
    \includegraphics[width=\textwidth]{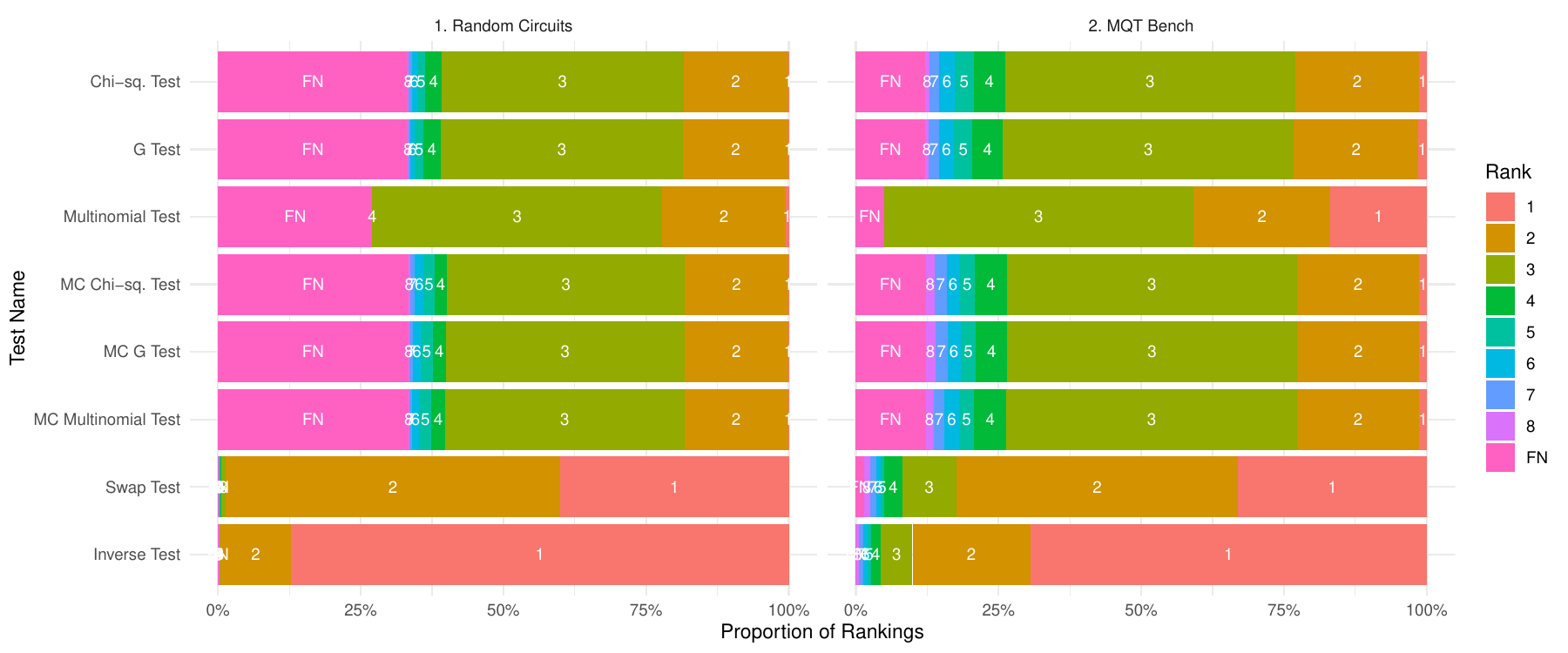}
  \caption{Ranking performance comparison.}
  \label{fig:ranking_dashboard}
\end{figure*}

\Cref{fig:rank_1_per_num_qubits} illustrates the change in the percentage of rank-1 outcomes across all tests and datasets. No clear trend emerges with respect to the number of qubits; however, additional datasets will be needed to determine whether this pattern is generalizable.

\anncmnt{\textbf{Answer to RQ1}: The \inverseTest performs best at reliably distinguishing between quantum states that are different.}

\begin{figure*}[tb]
  \centering
    \includegraphics[width=\textwidth]{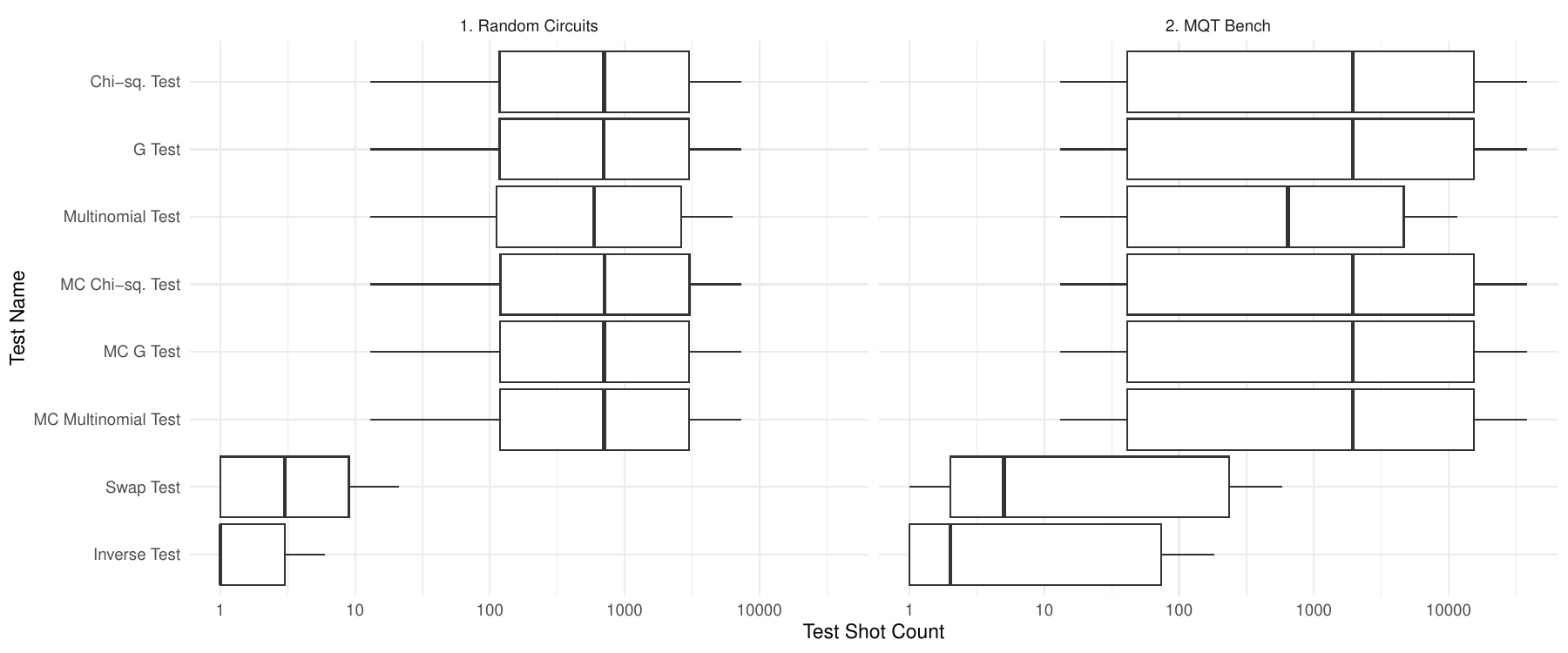}
  \caption{Shot count performance comparison (outliers not visualized). False negatives are excluded from the plots.}
  \label{fig:count_dashboard}
\end{figure*}
{\begin{table}[t]
\centering
\caption{Performance of each tests in our empirical study. Top-3 best values are reported in \textbf{bold}.}
\vspace{-1em}\resizebox{\columnwidth}{!}{%
\begin{tabular}{@{}l|rrr|rrr@{}}
\toprule
\textbf{Test Name} & \multicolumn{3}{c|}{\textbf{Random Circuits dataset}} & \multicolumn{3}{c}{\textbf{MQT Bench dataset}} \\
                   & \textbf{TP} & \textbf{FN} & \textbf{Recall} & \textbf{TP} & \textbf{FN} & \textbf{Recall} \\
\midrule
$\chi^2$ test           & 9.77E+07 & 7.75E+07 & 0.558 & 3.87E+06 & 6.28E+05 & 0.860 \\
G-test                 & 9.79E+07 & 7.73E+07 & 0.559 & 3.88E+06 & 6.25E+05 & 0.861 \\
Multinomial test       & 1.25E+08 & 5.04E+07 & 0.712 & 4.26E+06 & 2.40E+05 & 0.947 \\
MC $\chi^2$ test        & 9.74E+07 & 7.78E+07 & 0.556 & 3.87E+06 & 6.32E+05 & 0.860 \\
MC G-test              & 9.76E+07 & 7.76E+07 & 0.557 & 3.87E+06 & 6.32E+05 & 0.860 \\
MC Multinomial test    & 9.75E+07 & 7.77E+07 & 0.557 & 3.87E+06 & 6.32E+05 & 0.860 \\
\swapTest              & 1.44E+08 & 3.09E+07 & \textbf{0.824} & 4.31E+06 & 1.89E+05 & \textbf{0.958} \\
\statevectorTest & 1.75E+08 & 0.00E+00 & \textbf{1.000} & 4.50E+06 & 0.00E+00 & \textbf{1.000} \\
\inverseTest           & 1.58E+08 & 1.68E+07 & \textbf{0.904} & 4.39E+06 & 1.14E+05 & \textbf{0.975} \\
\bottomrule
\end{tabular}}
\label{tab:confusion_matrix_case_study}
\end{table}}

\smallskip To evaluate differences in shot count, \Cref{fig:count_dashboard} provides useful insights. For the MQT Bench dataset, the \textbf{\inverseTest requires significantly fewer shots}: the median is below 10 for both the \inverseTest and the \swapTest, with the former requiring slightly fewer shots (2 vs.\ 5). In contrast, the statistical tests have a median shot count $\approx$1900 (with the exception of the multinomial test with $\approx$600).

For the Random Circuits dataset in \Cref{fig:count_dashboard}, the difference is also pronounced: the median \textbf{number of shots for the \inverseTest is 3 and for the \swapTest is 9}. The statistical tests have a median between approximately 600 and 700 shots. 

We do not include outliers in \Cref{fig:count_dashboard} due to the large number of observations, which makes plotting them computationally expensive. The whiskers of the box-and-whisker plot indicate that the 1.5 interquartile range is relatively small: less than $10^4$ for Random Circuits and less than $10^5$ for the MQT Bench. However, it is important to note that some cases exceeded these values. In fact, the maximum number of shots observed in both datasets reached the upper limit allowed: approximately $10^5$ for Random Circuits and $10^7$ for the MQT bench. Thus, in practice, a large number of shots may still be required, especially when the extent of difference between the faulty and correct circuits is unknown.

\Cref{tab:confusion_matrix_case_study} further supports these findings. By construction, the \statevectorTest are independent of shot count and achieves perfect recall.  Among the remaining tests, recall is generally lower for the Random Circuits dataset than for the MQT Bench dataset. This is expected due to the difference in shot counts ($10^5$ vs.\ $10^7$), highlighting the importance of having sufficient data for reliable testing. Notably, the relative drop in recall is smallest for the \inverseTest ($0.071 \leftarrow 0.975 - 0.904$), followed by the \swapTest ($0.134 \leftarrow 0.958 - 0.824$), and then the statistical tests (e.g., the $\chi^2$ test shows a drop of $0.302 \leftarrow 0.860 - 0.558$).

For Random Circuits dataset, \textbf{the \inverseTest achieves the highest recall ($0.904$)}, followed by the \swapTest test ($0.824$) and then the statistical tests (approximately $0.56$ for all, except the multinomial test, which is prone to ``latching'' and reaches $0.712$). For the MQT Bench dataset, the \textbf{\inverseTest and \swapTest have similar recall values ($0.975$ and $0.958$, respectively)}, while most statistical tests perform lower (around $0.86$), with the multinomial test achieving a higher recall of $0.95$.

Overall, \textbf{the results suggest that quantum-centric tests (Inverse and Swap) provide more robust performance}. As a side note, our analysis indicates that about 3.3\% of cases expected to be flagged  were missed, slightly below the anticipated 5\% (as $P_e = 0.05$). This implies that the theoretical formula for required shot count from \Cref{eq:estimate_of_s}, aligns well with empirical results and may even be slightly conservative in practice.

\anncmnt{\textbf{Answer to RQ2}: In terms of the median, the \inverseTest requires 60\% to 66\% less shots than the second best, the \swapTest, to detect true positives while keeping false negatives low.}

\subsection{Threats to validity}
We classify validity threats as suggested by \citet{yin2009case,wohlin2012experimentation}.

\subsubsection{Internal validity}
We developed our own codebase, which may introduce implementation errors. To reduce this risk, we wrote unit tests, conducted code reviews, and followed standard practices like version control and documentation to ensure correctness and traceability.

\subsubsection{Construct validity}
To ensure validity, we evaluated each test under controlled conditions with known ground truth and modeled outcomes as binary classification problems, following standard practice in empirical software testing~\cite{arcuri2011practical, wohlin2012experimentation, sjoberg2022construct}. We used established statistical baselines and explicitly defined quantum unit test procedures to minimize ambiguity. 

\subsubsection{External validity}
Throughout the previous sections, we assume execution of the tests on a Fault-Tolerant (FT) quantum computer or an idealized, noise-free simulator. Although the same programs can be executed on Noisy Intermediate-Scale Quantum (NISQ) hardware using error mitigation techniques~\cite{takagi2022fundamental,cai2023quantum}, our focus remains on FT devices or simulators. This decision is motivated by the projection that FT quantum computers, expected to become available around 2029 (e.g., IBM promises a computer able to execute ``100 million quantum gates on 200 logical qubits~\cite{ibm2025ft}''), are the most likely to deliver quantum advantage and generate practical business value~\cite{campbell2017roads}. Thus, prioritizing the FT backend is both a forward-looking and a pragmatic approach.

Software engineering studies often face challenges due to the variability of real-world environments, and the issue of generalization remains inherently difficult to resolve~\cite{wieringa2015six}. Further threats to external validity stem from our design choices.

First, the quantum programs evaluated in this study are quantum algorithms which may or may not reflect the diversity of real-world quantum applications, potentially limiting generalizability.  To minimize this threat we also considered 10,000 randomly generated quantum circuits.

Second, the fault model is based on the predefined set of mutation operators proposed by \citet{QMutPyJournal,QMutPyToolShort,QMutPyTool} and \citet{Muskit}\footnote{Main difference between \citet{QMutPyJournal,QMutPyToolShort,QMutPyTool}'s mutation operators and \citet{Muskit}'s is that the former only inserts or replaces a gate to a syntactically equivalent one rather to any gate.  Although a more restrict approach, Fortunato \emph{et al.}'s do not generate invalid mutants by design.} and a percentage of the generated mutants. These constraints may bias the fault set toward simpler or faster-to-generate cases, underrepresenting faults that occur in practice.
However, the lack of large datasets of quantum faults that have occurred in real quantum software makes it impossible to truly assess this.
QBugs~\cite{QBugs} is not available.  The Bugs4Q~\cite{ZHAO2023111805} dataset of \emph{bugs} in Qiskit only contains 20 \emph{bugs} (out of 42) related to the source code of a quantum program, and most could be mimicked by the set of mutation operators we used in our study.  The remaining ones are, for example, related to refactors\footnote{Bugs4Q \#5: \url{https://github.com/Z-928/Bugs4Q-Framework/blob/main/qiskit/5/modify_5.txt}, accessed July 2025.}, related to the draw of the circuit\footnote{Bugs4Q \#22: \url{https://github.com/Z-928/Bugs4Q-Framework/blob/main/qiskit/22/modify_22.txt}, accessed July 2025.}, or related to the execution of the circuit\footnote{Bugs4Q \#4: \url{https://github.com/Z-928/Bugs4Q-Framework/blob/main/qiskit/4/modify_4.txt}, accessed July 2025.}; which are outside the scope of this paper.

Third, regarding the pairs programs-mutants, our set is similar to the one used by \citet{UsandizagaMutation2025}, for a number of qubits up to 5.
We considered 85 circuits from the MQT Bench and they considered 80 (the difference is due to the usage of a more recent version of the MQT Bench).  For the 85 circuits we generated 45,030 mutants vs.\ 45,067 (some either missed or with compilation issues)\footnote{\url{https://github.com/EnautMendi/Quantum-Circuit-Mutants-Empirical-Evaluation/issues/2}, accessed July 2025.}.

Finally, our evaluation uses a fixed set of statistical %
and $p$-value thresholds (i.e., 0.05 and 0.01).  Although our evaluation considers the statistical test commonly used in quantum software testing~\cite{wang2022qusbt,wang2023qucat,UsandizagaMutation2025}, i.e., the $\chi^2$ test, we also consider five other powerful tests.  Nevertheless, different statistical tests or parameterizations could yield different outcomes. These limitations highlight the need for future work exploring a broader corpus of quantum programs, fault models, and evaluation baselines.

Our theoretical complexity analysis is dataset-agnostic, which supports broad generalizability. However, some tactical questions (such as the relationship between test ranks and the number of qubits discussed in \Cref{sec:case_study_results}) may require further investigation. This opens opportunities for other researchers to replicate our work on alternative datasets and assess its generalization to different quantum software systems.

\section{Discussion}\label{sec:discussion}

In this section we briefly describe the novelty of our work, the three main takeaways, and finally discuss a practical approach to design a testable quantum circuit.

\subsection{Novelty}

\begin{itemize}[leftmargin=*]
  \item[\small{$\bigstar$}] The \statTests and the \statevectorTest we use follow established approaches commonly used in both the literature and practice. 

  \item[\small{$\bigstar$}] The \swapTest is well-known; we have made a minor modification to simplify classical post-processing.

  \item[\small{$\bigstar$}] As for the \inverseTest, the underlying concept (using a reversible circuit~\cite{1317002,6231097,7977062,1197682,syamala2012,patel2021qraft,MondalReversible2022}) is standard in quantum programming. For example, it is commonly applied in NISQ devices for noise assessment. However, to the best of our knowledge, our formulation is novel in the context of quantum unit tests.

  \item[\small{$\bigstar$}] Regarding the efficiency and effectiveness of these tests, we also have not seen any comparison.
\end{itemize}

\subsection{Takeaways}

\begin{enumerate}[leftmargin=*]%
  \item By construction, both quantum-centric tests, the Inverse and Swap tests, yield zero false positives (see \Cref{sec:methodology_swap_test,sec:methodology_quite_measurement_test}).

  \item The \inverseTest performs best, followed closely by the \swapTest, at reliably distinguishing between quantum states that are slightly different (see \Cref{sec:case_study_results}).  The \statevectorTest also shows excellent results, with zero false positives and false negatives; however, it is not scalable (see \Cref{sec:methodology_quite_statevector_test}).

  \item %
  In theory, classical resource requirements depend on the setup and how the actual and expected states are represented (see \Cref{sec:complexity_analysis}).  In practice,
  the \inverseTest requires the fewest shots (potentially translating to lower quantum compute time), followed by the \swapTest (see \Cref{sec:case_study_results}). \\
  While statistical tests are generally the cheapest in terms of raw compute cost, their long-term expense may be higher due to the overhead of handling false positives and false negatives, which often demands significant human effort.
\end{enumerate}

\subsection{Designing Testable Quantum Circuits}\label{sec:design}

To ensure that the software is testable, we can design Python functions that generate quantum circuits in a scalable manner, based on an input parameter $n$ (e.g., consider the \texttt{QFT} Python class\footnote{This class generates a circuit for the Quantum Fourier Transform~\cite{coppersmith2002approximate}.} in Qiskit~\cite{qft_qiskit}, which accepts \texttt{num\_qubits} as an input parameter).

These functions dynamically adjust the circuit's structure according to $n$. The implementation typically involves control flow constructs, such as loops and conditional branches, that depend on $n$. Therefore, we can apply standard software testing practices, such as path coverage (where every code path is executed at least once) and control structure testing, which includes simple and nested loops.

Which values of $n$ should we choose for testing? Here, we can apply the \textit{equivalence partitioning} approach to select representative values. The partitions depend on the subroutine being tested, but the following cases are commonly relevant. 
\begin{itemize}[leftmargin=*]
    \item For $n = 1$, the circuit may consist only of single-qubit gates (although this is not always the case).
    \item For $n = 2$, two-qubit gates (e.g., \cnot) may be introduced.  
    \item For $n = 3$, multi-controlled gates like the Toffoli (\mcx{2}, where $2$ represents the number of control qubits) gate may appear.  
    \item For $n \geq 4$, more general multi-controlled gate (\mcx{n-1}) may be required.  
\end{itemize}

While we can create tests for large values of $n$, this may conceptually contradict the nature of a unit test, where we aim to construct simple, isolated inputs for which it is easy to compute the expected output. The primary goal of unit testing is to cover potential execution paths within the subroutine under test while maintaining clarity and feasibility in verifying expected results. Therefore, we suggest designing subroutines in a way that all execution paths can be covered, if possible, by testing the inputs for small values of $n$. 

\paragraph*{Example} Suppose that we aim to test Grover's algorithm~\cite{grover1996fast} and our $n=128$, a size that cannot be simulated classically (as the state vector has $2^{128}$ elements). Assuming that the implementation of the algorithm follows the principles described above, we can create smaller versions of circuits for $n=1, 2, 3, 4$ (using various values of $W$). This approach enables reasonably thorough testing of the code in a simulator while keeping computational costs manageable.

\section{Conclusions and Future Work}\label{sec:conclusion}

\paragraph*{Conclusion} 

In this paper, we present an evaluation of four representative test groups: Statistical, Statevector, Swap, and the novel Inverse test. The evaluation is based on controlled experiments conducted on both synthetic and benchmark quantum circuits.

Our findings indicate that, although statistical tests remain standard in many contexts, their destructive nature and reliance on high shot counts limit their practicality. The \statevectorTest, which requires complete state information, is infeasible on real quantum hardware. The \swapTest, which operates on an auxiliary qubit without altering the data qubits state (if the test passes), produced low false positive rates in our experiments but showed lower recall than the \inverseTest.

The proposed \inverseTest demonstrates high statistical power, no false positives, and strong recall across all tested datasets. It also requires lower number of shots than the other tests in many cases. While its worst-case circuit preparation complexity is theoretically exponential, this is often not a limiting factor in practice. As a result, the \inverseTest offers a computationally efficient and effective alternative. In light of our findings, we encourage the community working on fault detection in quantum circuits to adopt the \inverseTest.

\paragraph*{Future work} Future work will further investigate the generalizability of these findings, beginning with input state variation. Specifically, we plan to extend our evaluation to circuits initialized with a broader range of inputs, including automatically generated states, as explored in prior work ~\cite{wang2023qucat,wang2022qusbt,wang2021quito,paltenghi2023morphq,xia2024fuzz4all,QuraTest}.
 We also aim to scale testing to larger and noisier circuits, integrate these methods into hybrid quantum-classical workflows, and develop composable testing toolkits to support practical quantum software engineering.

\section*{Acknowledgments}
The authors thank profusely Shaukat Ali and Paolo Arcaini for their insightful discussions and valuable contributions to this research. Our heartfelt thanks also go to the organizers of the Dagstuhl Seminar 24512\footnote{Dagstuhl Seminar 24512 -- Quantum Software Engineering homepage: \url{https://www.dagstuhl.de/en/seminars/seminar-calendar/seminar-details/24512}, accessed July 2025.} namely, Shaukat Ali, Johanna Barzen, Andrea Delgado, Hausi A. M\"{u}ller, and Juan Manuel Murillo, for bringing us together and catalyzing our collaboration.

This work was partially supported by the Natural Sciences and Engineering Research Council of Canada (grant \# RGPIN-2022-03886), the LASIGE Research Unit, ref.\ UID/00408/2025 - LASIGE, the QCloud QuantumEd project funded by the EOSC \textit{INFRAEOSC-03-2020} (grant \#101017536),  CyberSkills HCI Pillar 3 Project 18364682, Science Foundation Ireland grant 13/RC/2094\_P2, and Q-SERV-Q\&T Project (PID2021-124054OB-C32, of the Ministry of Economy, Industry and Competitiveness and FEDER). The authors thank the Digital Research Alliance of Canada for providing computational resources.

\printbibliography

\end{document}